\documentclass[11pt,preprint]{aastex}
\pdfoutput=1
\usepackage{graphicx}

\let\oldfootsep=\footnotesep
\newcommand\ltsima{$\; \buildrel <\over\sim \;$}
\newcommand\simlt{\lower.5ex\hbox{\ltsima}}
\newcommand\gtsima{$\; \buildrel >\over\sim \;$}
\newcommand\simgt{\lower.5ex\hbox{\gtsima}}

\newcommand\msun {M_\odot}

\newcommand\mearth {{M_\oplus}}

\newcommand\umin{u_{\rm min}}

\newcommand\pac{Paczy{\'n}ski }
\newcommand\ie{{i.e.}}

\newcommand\rep {\tilde{r}_E}

\shorttitle{}
\shortauthors{Bennett et al}

\begin{document}


\title{Planetary and Other Short Binary Microlensing Events\\ from the MOA Short Event Analysis}


\author{D.P.~Bennett\altaffilmark{1},
T.~Sumi\altaffilmark{2}, 
I.A.~Bond\altaffilmark{3}, 
K.~Kamiya\altaffilmark{4}, \\
F.~Abe\altaffilmark{4}, 
C.S.~Botzler\altaffilmark{5}, 
A.~Fukui\altaffilmark{4}, 
K.~Furusawa\altaffilmark{4}, 
Y.~Itow\altaffilmark{4},  
A.V.~Korpela\altaffilmark{6},
P.M.~Kilmartin\altaffilmark{7}, 
C.H..~Ling\altaffilmark{3},
K.~Masuda\altaffilmark{4},  
Y.~Matsubara\altaffilmark{4}, 
N.~Miyake\altaffilmark{4}, 
Y.~Muraki\altaffilmark{4}, 
K.~Ohnishi\altaffilmark{8}, 
N.J.~Rattenbury\altaffilmark{5}, 
To.~Saito\altaffilmark{9},
D.J.~Sullivan\altaffilmark{6}, 
D.~Suzuki\altaffilmark{2}, 
W.L.~Sweatman\altaffilmark{3},
P.J.~Tristram\altaffilmark{7}, 
K.~Wada\altaffilmark{2}, and
P.C.M.~Yock\altaffilmark{5} \\
(The MOA Collaboration)\\ \ \\
              } 



\keywords{gravitational lensing: micro, planetary systems}

\altaffiltext{1}{Department of Physics,
    University of Notre Dame, Notre Dame, IN 46556, USA; 
    Email: {\tt bennett@nd.edu}}
\altaffiltext{2}{Department of Earth and Space Science, Osaka University, Osaka 560-0043, Japan}
\altaffiltext{3}{Institute for Information and Mathematical Sciences, Massey University, Auckland 1330, New Zealand}
\altaffiltext{4}{Solar-Terrestrial Environment Laboratory, Nagoya University, Nagoya, 464-8601, Japan}
\altaffiltext{5}{Department of Physics, University of Auckland, Private Bag 92-019, Auckland 1001, New Zealand}
\altaffiltext{6}{School of Chemical and Physical Sciences, Victoria University, Wellington, New Zealand}
\altaffiltext{7}{Mt. John Observatory, P.O. Box 56, Lake Tekapo 8770, New Zealand}
\altaffiltext{8}{Nagano National College of Technology, Nagano 381-8550, Japan}
\altaffiltext{9}{Tokyo Metropolitan College of Aeronautics, Tokyo 116-8523, Japan}

\clearpage

\begin{abstract}
We present the analysis of four candidate short duration
binary microlensing events from the 2006-2007 MOA Project
short event analysis. These events were discovered as a
byproduct of an analysis designed
to find short timescale single lens events that may be due to free-floating
planets. Three of these events are determined to be microlensing events, while
the fourth is most likely caused by stellar variability.
For each of the three microlensing events, the signal is almost entirely due
to a brief caustic feature with little or no lensing attributable mainly to the
lens primary. One of these events, MOA-bin-1, is due to a planet, and it is the
first example of a planetary event in which stellar host is only detected
through binary microlensing effects. The mass ratio and separation are
$q = 4.9\pm 1.4\times 10^{-3}$ and $s = 2.10\pm 0.05$, respectively. A
Bayesian analysis based on a standard Galactic model indicates that
the planet, MOA-bin-1Lb, has a mass of $m_p = 3.7\pm 2.1 M_{\rm Jup}$, and orbits a star
of $M_\ast = 0.75{+0.33\atop -0.41}\msun$ at a semi-major axis of 
$a = 8.3 {+4.5\atop -2.7}\,$AU. This
is one of the most massive and widest separation planets found by microlensing. 
The scarcity of such wide separation
planets also has implications for interpretation of the isolated planetary
mass objects found by this analysis. If we assume that we have been
able to detect wide separation planets with a
efficiency at least as high as that for isolated planets, then we can
set limits on the distribution on planets in wide orbits. In particular, if
the entire isolated planet sample found by \citet{sumi11}
consists of planets bound in wide orbits around stars, we find that it is 
likely that the median orbital semi-major axis is $> 30\,$AU.
\end{abstract}
\clearpage


\section{Introduction}
\label{sec-intro}
Gravitational microlensing plays a unique role among methods to search for 
exoplanets because planets can be detected without the detection of any
light from the host star \citep{bennett_rev,gaudi_rev}. In fact, old, planetary 
mass objects can be detected even in cases when there is no indication of
a host star \citep{sumi11}. This contrasts with the Doppler radial velocity
\citep{butler-catalog,bonfils11} and transit \citep{borucki11} methods, which rely
on the precision measurements of light from the source star to the
signals needed to detect the planets. 

Until 2011, the primary implication of the microlensing searches
for exoplanets was the relatively large number of cold planets
orbiting sub-solar-mass stars \citep{sumi10,gould10,cassan12}.
Microlensing is primarily sensitive to planets beyond the snow line,
where the proto-planetary disk is cold enough for ice to condense.
The snow line 
\citep{ida05,lecar_snowline,kennedy-searth,kennedy_snowline,thommes08}
is of great importance in the core accretion theory. Ices can 
condense beyond the snow line, and this increases the density of solid material
in the proto-planetary disk by a factor of a few, compared to the
disk just inside the snow line. This high density of
solids beyond the snow line is needed to form the ice-rock cores that
can grow massive enough to accrete significant amounts of hydrogen
and helium gas to become gas giants. But, it is also thought that 
this process may often be terminated before the final gas accretion
phase, particularly for low-mass stellar hosts \citep{laughlin04}, This
leaves a large number of $\sim 10\mearth$ ``failed jupiter" planets,
which are consistent with the microlensing observations.

A somewhat more unexpected discovery was reported by
\citet{sumi11} who have analyzed the first two years of the MOA-II Galactic
bulge microlensing survey data, and found an excess of events
with an Einstein radius crossing time of $t_E < 2\,$days. This excess
could not be explained by an extrapolation of the brown dwarf mass function
into the planetary mass regime, and so a previously undetected 
population of isolated planetary mass objects was needed to explain it. Such a
population is expected to arise from a variety of processes, such as
planet-planet scattering \citep{levison98,ford08,guillochon11}, star-planet scattering 
\citep{holman99,musielak05,doolin11,malmberg11,veras12}, and
stellar death \citep{veras11}. However, \citet{sumi11} find
that this new population consists of $1.8{+1.7\atop -0.8}$ 
times as many $\sim 1$ Jupiter-mass objects as main sequence stars,
and this is somewhat larger than expected. But, as
\citet{sumi11} show, the 
number of bound planets found by microlensing is similar to the number
of planets in the new isolated planet population.

The planetary mass objects in this newly discovered population
are isolated in the sense that there is no host star that can be detected
in the microlensing data. This implies a lower limit on the projected
separation between the planet and a possible host star that depends on
the hypothetical host star's Einstein radius. The lower limits reported
by \citet{sumi11} range from 2.4 to $15.0\,R_E$, which corresponds to
7-$45\,$AU, assuming typical host star masses and random orientations.
The fact that distant host star are not strongly excluded
has led some \citep{nagasawa11,wamb11} to suggest that many of 
these isolated planets may, in fact, be bound in distant orbits to host stars.
Others have even suggested that the large number of isolated planets
may be due to an {\it in situ} formation mechanism \citep{bowler11}
that differs from the core accretion theory \citep{lissauer_araa}.

It remains uncertain, however, if most of the planets in this newly
discovered isolated planet population are truly unbound, or orbiting
stars in very wide orbits. And if a significant fraction of these
isolated planets do orbit stars, it is unclear if they might have
formed somewhat closer to their host stars (near the snow line)
and moved outwards since their formation. Alternatively, distant
bound planets could have formed {\it in situ} by gravitational instability
\citep{boss97}.

These questions can be addressed with microlensing survey data
by identifying microlensing events due to planets in very wide orbits.
When we are lucky enough to have the relative motion of the lens
and source stars align with the planet-star separation vector, then
it is possible to detect the star-planet system as
two single lens light curves that are widely separated in time
\citep{distefano99}. But, if the planet-star separation is not too large,
it is also possible to detect the binary lens effects of the distant 
host star through a close approach of the source star with the planetary
caustic \citep{han_widepl2003,han_widepl2005}. 
These appear as short duration binary microlensing events,
but not all short duration binary lensing events are due to low-mass
lens systems. Stellar binary lenses will generate small caustics if
their separation is much smaller than the Einstein radius. It is
also possible to have a short caustic crossing feature in a much longer
duration microlensing event, and if the source is faint, it may be 
difficult to detect the microlensing magnification outside of the
caustic if the source is blended with brighter stars. So, detailed
modeling of the observed binary events is needed to determine
which of these scenarios provides the correct description of a 
given microlensing event.

The main background for short duration microlensing events is
variable stars with a short duration, such as cataclysmic variables
(CVs) or flare stars, as discussed by \citet{sumi11}. Such variables
are much more common than short duration microlensing
events, but they can generally be distinguished from microlensing events
because their light curves are poorly fit with microlensing models.
However, it is more difficult to distinguish between variable stars and
binary microlensing events because the binary events are described
by 3 or 4 additional parameters. If there are a small
number of data points sampling the important light curve features, it will 
be much easier to fit a variable star light curve with a binary
lens model than with a single lens model.

In this paper, we present the analysis of four light curves
from the analysis of \citet{sumi11}. These light curves all exhibit
short timescale brightenings that can be well fit by binary microlensing
light curves. Following \citet{bond04}, we define events with a
mass ratio of $q \leq 0.03 $ as planetary events, while a mass
ratio of $q > 0.03$ would imply a brown dwarf secondary.
We find that one of these events, MOA-bin-1, is due to
microlensing by a wide-separation planet, which is separated from its
host star by more than two Einstein radii. Two of the events, 
MOA-bin-2 and MOA-bin-3, are best fit with binary models 
with brown dwarf mass ratio secondaries. The final event has
a light curve that is best fit by a primary lens of about a Jupiter mass,
orbited by a secondary of a few Earth-masses. However, the brightening
is extremely short, and the apparent binary lens features are just
barely sampled by the observations. Thus, they may be fit by a binary
lens model because such a model has enough adjustable parameters
to fit the observations. Unlike the case of MOA-bin-1, the apparent 
microlensing light curve features are not oversampled, so the
observations do not definitively demonstrate that the features are
due to microlensing. It seems most likely that, despite the
good microlensing fit, the brightening is due to a large amplitude
stellar flare.

This paper is organized as follows. In Section~\ref{sec-data}, we describe
the data analysis and the selection of events to analyze in this paper.
Section~\ref{sec-lc} presents the details of the light curve analysis for
each of the four events that we consider, and we present our analysis
of the physical properties of the MOA-bin-1L planetary system
in Section~\ref{sec-prop}. We discuss the implications
of these and similar events in resolving the puzzles raised
by our recent discovery \citep{sumi11} of a large population of 
isolated planetary mass lens masses in Section~\ref{sec-wide},
and we summarize  our conclusions
in Section~\ref{sec-conclude}. In Appendix~\ref{sec-moa-bin1-nopl},
we further discuss the rationale for excluding non-planetary
models for MOA-bin-1, and finally, in the Appendix~\ref{sec-src_col},
we present an analysis of the
source star colors for two of the three binary lensing events presented
in this paper, plus the 10 isolated planet events from 
\citep{sumi11}. For five of these events the Optical Gravitational 
Lensing Experiment (OGLE) data allows us to
use an improved version of the MOA-OGLE color method
introduced by \citet{gould_col}. This analysis shows that
the source stars are likely to be normal bulge main sequence or sub-giant
stars, although in two of these five cases, the color of the source star is
difficult to determine due to microlensing model uncertainties.

\section{Data Analysis and Event Selection}
\label{sec-data}

The light curve analysis presented here is based on the data analysis and
preliminary event selection of \citet{sumi11},
which is discussed in some detail in the Supplementary Information for that
paper. This analysis uses the 2006-2007 MOA-II galactic bulge survey data, which
consist of $\sim 8250$ images
of each of the two most densely sampled fields (fields gb5 and gb9) and
1660-2980 images of each of the 20 other less densely sampled fields. 
The bulge was observed for about 8 months of each year from the end of 
February to the beginning of November. This observing strategy is designed to
detect short duration single lens events and short timescale 
anomalies in the light curves of stellar microlensing events
due to planets orbiting the lens stars
\citep{mao91,bennett_rev,gaudi_rev}.
The MOA images were reduced with MOA's implementation \citep{bond01}
of the difference image analysis (DIA) method \citep{tom96,ala98}. 

The images were taken using the custom MOA-Red wide-band filter, which
is approximately equivalent to the sum of the standard Kron/Cousins $R$ and $I$-bands, and
the photometry was approximately calibrated to the Kron/Cousins $I$-band using OGLE-II 
photometry map of the Galactic bulge \citep{uda02}, as described by \citet{sumi11}.
In this paper we improve upon this calibration with a more careful comparison to
the OGLE-III photometry catalog \citep{ogle3-phot}.

Candidate short duration binary microlensing
events were selected following the procedure of \citet{sumi11}, using 
cuts 0 and 1 to identify select events. The 5-parameter microlensing fits used
for \citet{sumi11} cut-2 were used to help identify short duration events, but these
5-parameter fits can sometimes fail to identify short duration events due to a well
known degeneracy \citep{alard97,distefano95} between short events with bright sources
and long events with faint sources.
Therefore, we also measured event durations with a simple top-hat filter that 
uses the duration of consecutive measurements that are $>3\sigma$ above the baseline.
All the events with at least 10 observations $>3\sigma$ above the baseline and
best fit single lens $t_E$ values or top-hat filter durations of
$ < 3\,$days were examined by eye in addition to all events with best fit single lens
events with $t_E < 3\,$days. The sample examined by eye consisted of 149
events, and this sample overlapped with 8421 events of all
durations classified by eye during the \citet{sumi11} analysis. About 40 of these
events were fit with microlensing models. Most of the events that were well fit by
microlensing models were well fit by single lens models. These included the 10 single
lens events of \citet{sumi11}, and a similar number of other events that were well fit
by single lens models, but did not pass all of the \citet{sumi11} selection criteria
(usually due to poor light curve coverage or blending uncertainties). The 18 events
from this sub-sample that were not well fit by single lens models, were then fit by
binary lens models. Most of these events could not be fit with a binary lens model, and they are
likely to be cataclysmic variables (CVs) with sharp changes the light curve slope that
are superficially similar to caustic entry features. A few other events were too
poorly sampled to yield a unique solution. This left the 4 events with good, short
duration binary microlensing models that we present in this paper.
The coordinates of these
events are listed in Table~\ref{tab-coords}.

Although this procedure is somewhat subjective, it is unlikely that 
any real, short duration, binary microlensing events were missed, as
long as they passed cuts 0 and 1 of \citet{sumi11}. The primary
backgrounds for the identification of short binary microlensing events are
short single-lens events and CVs. But, single lens events can all be
identified by their single lens fits (which were already done by \citet{sumi11}.
CV light curves typically look very different from microlensing event
light curves, and we have fit all the light curves that have a remote
possibility of being due to microlensing. So, it is unlikely that any
binary microlensing events have been missed by this procedure. The
one thing missing from this analysis is that we have not determined
the detection efficiency for these short binary lens events.

As in the case of the \citet{sumi11} analysis, once the events were identified,
we requested contemporaneous OGLE-III data from the OGLE collaboration.
Unfortunately, for 3 of the 4 light curves we present, there is no OGLE-III data
for the star in question during the year that the event occurred. And for MOA-bin-1,
the event with OGLE-III, coverage, the OGLE observations are limited to times of
modest magnification. The lack of signal in the OGLE data makes it difficult to 
identify the location of the lensed source in the OGLE images, particularly when
the images with the most significant magnification have poor seeing, as is the
case for MOA-bin-1. In this case, OGLE have reported the photometry 
at the location of an apparent star identified in their reference frames,
which is 2-3 times brighter than the source star, so it is possible that 
the source location is significantly offset from the location used for
the photometry. This could lead to systematic errors in the OGLE 
photometry for this event.

\subsection{Correction for Systematic Photometry Errors}
\label{sec-syserr}

Two of these events had evidence for systematic photometry errors in the 
original photometry of the event. As is generally the case in microlensing
analyses, these systematic errors are not large enough to have an effect
on the event selection. Instead they affect the interpretation of individual
events.

The photometry of MOA-bin-3 originally
showed systematic nightly variations that were correlated with seeing, and
different reductions favored a close-binary, brown-dwarf secondary model
or a wide binary, planetary secondary model. Both models can produce
light curve shapes that are very similar during the times
that are sampled by the MOA data. So, it was worthwhile to see if
photometric re-reductions could improve the photometry of this event.
The MOA images for this event were re-reduced several times with
different choices for the reference images. These re-reductions
used the so-called ``cameo" images, which are sub-frames of the
original images, centered on the location of the event.
We selected a cameo data set which had a negligible correlation
between the seeing and the photometry in the unmagnified part
of the light curve.

The original MOA photometry for MOA-bin-1 also showed evidence of
systematic photometry errors. A telltale sign of the problem for this event
was a systematic pattern of residuals in the difference images taken at
high airmass. The direction of this systematic residuals varied with
the observing season, which implies that they are likely to be caused
by differential refraction. Such photometry problem could be caused by
the effect of differential refraction on bright reference stars that are
used to register the images if these stars have different colors.
Similar photometry errors might also arise if the target star is blended
with a star with a different color. The effect of these photometry errors
could also be seen in the binned MOA light curves as a slow decrease
in brightness for each season, which continued in the later seasons
that weren't included in this analysis.

To correct for this systematic error, we fit the photometry outside of the
microlensing event with a general quadratic function depending on
$(\cos \psi /\cos z)$ and $(\sin\psi /\cos z)$, where $z$ is the angle
between the target position and zenith and $\psi$ is the direction of 
differential refraction. ($1/\cos z =\,$airmass.) This quadratic model
is then used to correct the photometry, resulting in an improvement
$\Delta\chi^2 = 789.7$ for the 7232 MOA observations, implying an
average improvement of $\sigma/3$. The systematic photometry
trend over each season is largely removed by this correction.
Such a low-level systematic error does not have any significant
effect on the shape of the cusp crossing feature that led to 
selection of this event, but it does effect the lower level variation
of the light curve before and after the cusp crossing, which are
primarily due to the stellar host star. Thus, correction of this
systematic error was crucial for determining the properties of 
the host star for this planet.

\section{Light Curve Analysis}
\label{sec-lc}

The light curve modeling for this paper was done with the modeling code of
\citet{bennett-himag} with a few modifications. First, the light curve calculations
now make use of the hexadecapole approximation \citep{gould-hex,pej_hey} in situations
when finite source effects are important but not extremely strong. Full numerical
integration is used whenever there is a caustic crossing, and when there is no
caustic crossing, the hexadecapole approximation is used when the difference 
between both the hexadecapole  and quadrapole and the quadrapole and point
source approximations are relatively small. When the difference between one pair
of these lower order approximations becomes too large, the 
full numerical integration as described in \citet{bennett-himag} is used.

An additional modification is needed to deal with events that are dominated
by caustic features which are far from the center of mass. Such events can
approach the Chang-Refsdahl approximation \citep{chang-refsdal79,chang-refsdal84},
where the gravitational field encountered by the light rays from the source is dominated by
a single mass plus the gradient of a gravitational field from a more distant mass. This
gradient field is parameterized by the shear $\gamma$, which is given by
$\gamma = 1/d_{\rm dist}^2$, where $d_{\rm dist}$ is the separation of the distant mass 
measured in units of its Einstein radius. 
This shear approximation has been used for theoretical
studies of extrasolar planets detected by microlensing \citep{gouldloeb92}, and this is
useful for a theoretical understanding of what parameters can be constrained in 
events like this. If we instead express the shear in units of the nearby mass
it becomes $\gamma = q/d_{\rm near}^2$, where $q$ is the mass ratio of the distant
to the nearby mass, and $d_{\rm near}$ is the separation in units of the Einstein radius of the
nearby mass. This holds because the Einstein radii for the distant and
nearby masses scales as $\sim \sqrt{q}$. Thus, models with similar
values of $\gamma$, but very different values of $d_{\rm near}$ and $q$ will give very
similar light curve shapes. So, the shear approximation is a good guide to
indicate possible light curve degeneracies, but 
this approximation is not accurate enough to use for
light curve modeling and does not reduce the need for time consuming
numerical calculations.

The main difficulty with using general purpose light curve modeling codes
for modeling the events we present in this paper is
that it is awkward to us use polar 
coordinates centered far from the location of lens features giving rise to the
observed magnification. The region of parameter space that must be sampled
is the region of significant magnification, and this is not located at the 
center-of-mass for the system or at the locations of either of the lens masses.
Instead, the locations of significant magnification are located at the positions
of the caustics.
So, for most of the modeling used in this paper, we
define the coordinate center to be the centroid of the caustic curve, which is 
approached or crossed by the source. This centroid is defined numerically
from a set of caustic points that have been transformed, using the lens 
equation from a set of points that are equally spaced on the critical curve in
the image plane. Once in the general vicinity of the solution, however, the
use of the parameter correlation matrix with the Metropolis algorithm,
as described by \citet{bennett-himag} is sufficient to deal with the
parameter correlations in the center-of-mass coordinate system.

This caustic centered modeling approach is not entirely general. First, since the
caustic locations are only known numerically, the centroid of a caustic will
have some dependence on numerical parameters. But, a more serious issue
occurs when the lens parameters change in such a way that the caustics
would merge. This would cause a sudden, discontinuous change in the location
of the center of the coordinate system, that would likely cause the fitting code to fail.
So, this code is only used in situations when reconnection of the caustics is 
unlikely, and modeling for this paper has generally been performed in parallel
using the caustic centered approach in parallel with the standard center-of-mass
centered approach.

For each of the four events presented in this paper, we have done a systematic
search for binary lens models following the method of \citet{bennett-himag}, but
using the caustic-centered coordinate system. For events MOA-bin-1 and 
MOA-bin-2, which are dominated by a single cusp crossing and caustic crossing
feature, respectively, we also did a search based on these observed features.
For MOA-bin-1, we investigated models in which the source crossed each
of the caustics of a binary lens system, including wide and close binaries,
as well as the so-called resonant systems, which have only a single connected
caustic curve. For MOA-bin-2, we explicitly considered all geometries of
caustic crossing binaries. In both cases, the results were identical to the 
results from the initial condition grid search method of \citet{bennett-himag},
using the caustic-centered coordinate system.

\subsection{MOA-bin-1}
\label{sec-moa-bin1}

\begin{figure}
\plotone{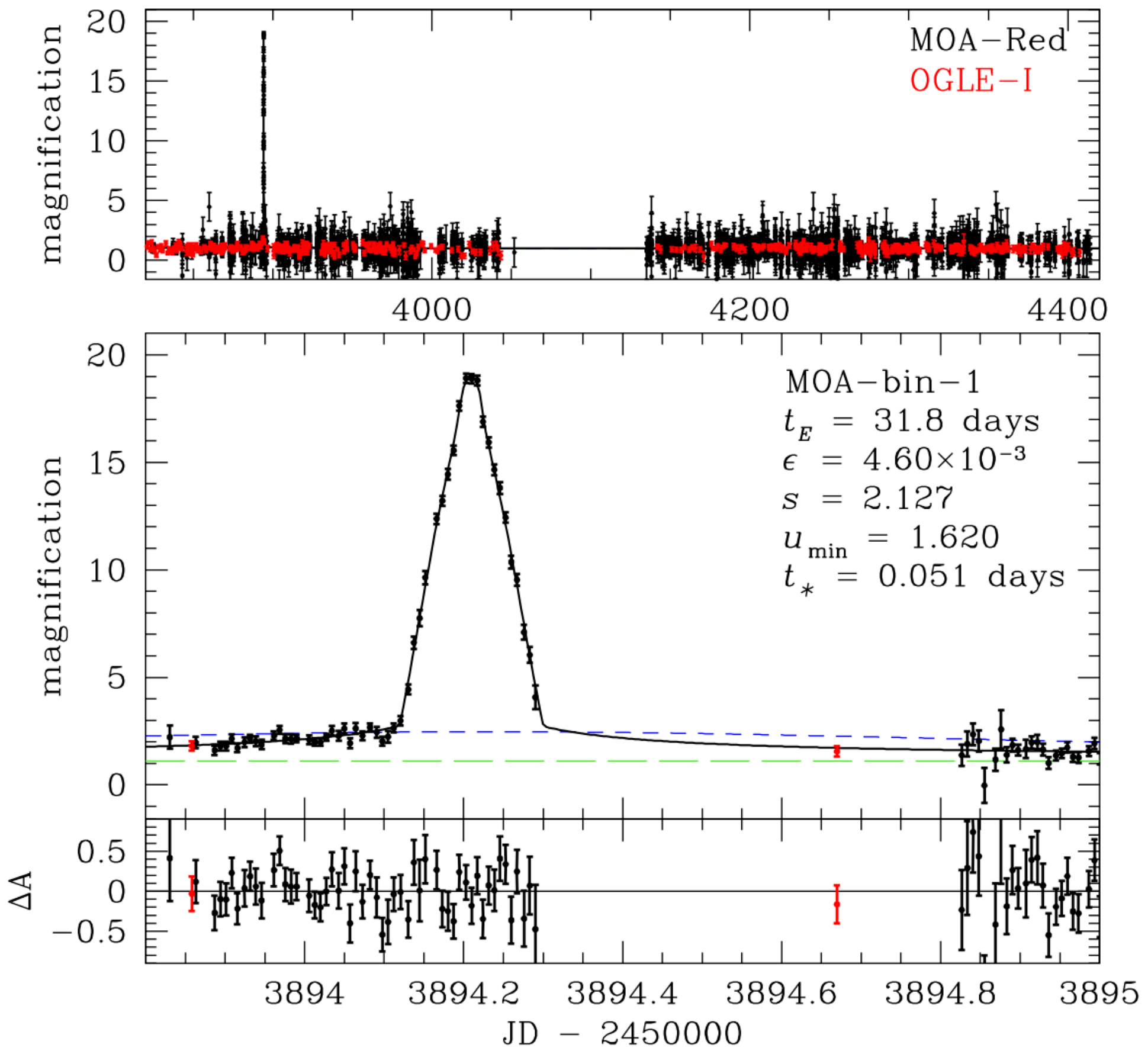}
\caption{The light curve of event MOA-bin-1 in field gb5 with a $\sim 10\,$min
sampling cadence. The top panel shows the 2006-2007
MOA-red and OGLE I-band data; the middle panel shows a close up
of the light curve peak, and the bottom panel shows the residuals
from the best fit. The blue, short-dashed curve and green, long-dashed
curves show the single lens light curves due to planet and host star, 
respectively. As described in the text, the OGLE data are not used in
the fit, but are added to the plot based on the cross-calibration of 
the OGLE and MOA photometry.
\label{fig-lc_bin1}}
\end{figure}

The light curve of the planetary event MOA-bin-1 is shown in Fig.~\ref{fig-lc_bin1}. This
event is in the gb5 field, which had a $\sim 10\,$minute sampling cadence, and it
has an inverted-V cusp crossing feature that is almost 
completely resolved by the MOA observations. Such a feature is readily 
recognized as a characteristic microlensing feature due to a caustic
crossing very close to a cusp, such that the separation of the caustic 
crossings along the source trajectory is similar to or smaller than
the source star diameter. (The brief, $\sim 20\,$minute duration flat peak
of the light curve indicates that the width of the caustic is just slightly
larger than the source.)

Unfortunately, the bright MOA-bin-1 cusp crossing feature occurred during daylight
hours at the OGLE telescope, and the OGLE data contain only three
relatively poor seeing observations with a magnification as large as
$A > 1.5$. These were not sufficient to identify the centroid of the 
of the event, so the photometry is done at the location of a ``star"
in the OGLE reference frame. This means that there are likely to be
blending-related systematic errors in the OGLE photometry.
Therefore, we do not include the OGLE data in our fits, but we
plot the OGLE data in Fig.~\ref{fig-lc_bin1} according to the 
color transformation discussed in Section~\ref{sec-prop}, assuming
a source star color consistent with an Hubble Space Telescope
(HST) color magnitude diagram (CMD),
as discussed in \citet{bennett08}. This puts the photometry from the relatively 
poor seeing OGLE data near the peak of the event on the curve, as indicated
in Fig.~\ref{fig-lc_bin1}, but the photometry from better seeing images at
lower magnification does not fit so well due to the offset between the 
location of the ``star" seen in the OGLE images and the location of 
the source star.

\begin{figure}
\plotone{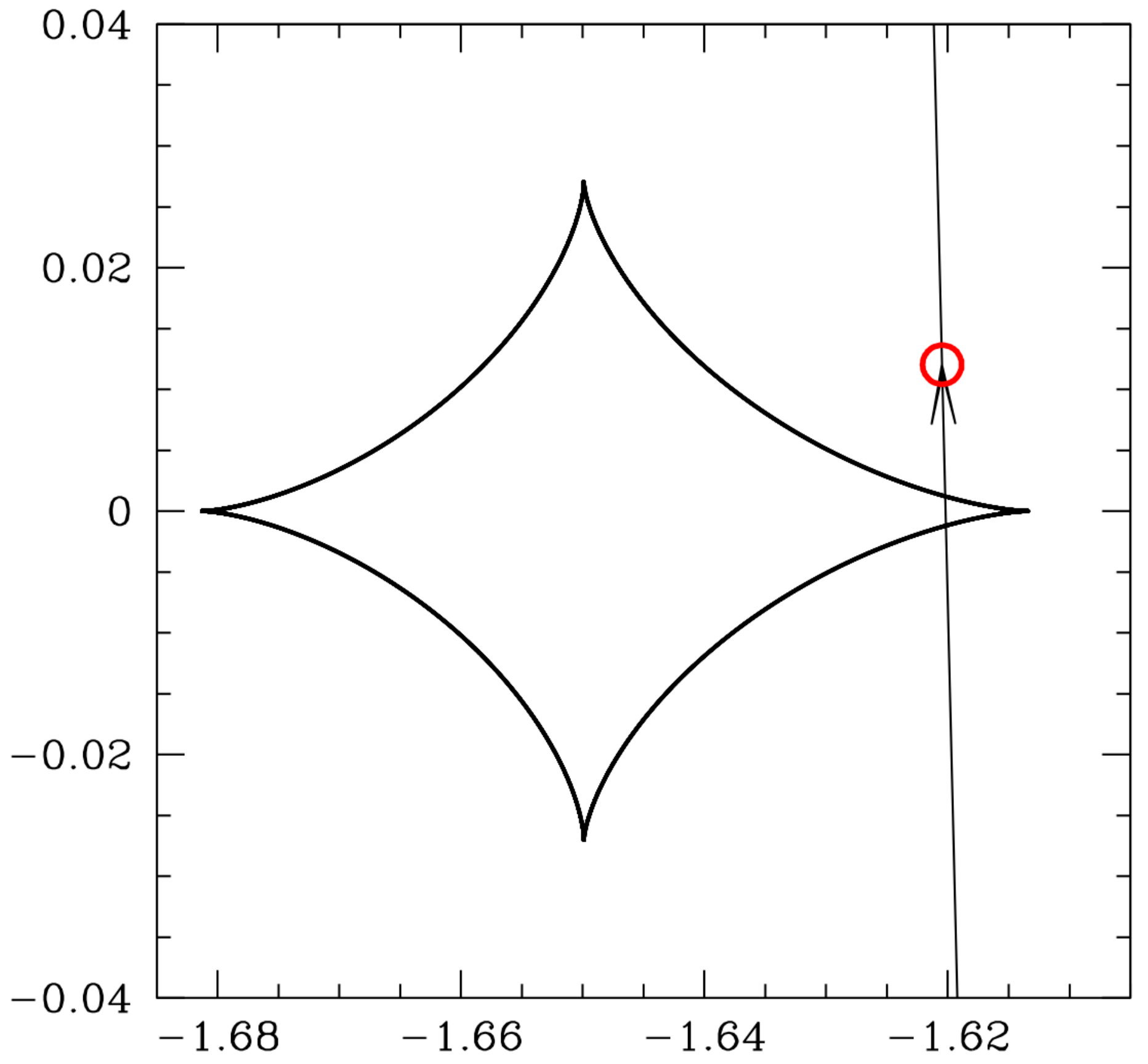}
\caption{ The caustic geometry for the best fit MOA-bin-1 model. The red circle indicates
the best fit source size, and the coordinates for this plot are centered at the 
center-of-mass of the full lens system (both the planet and its host star). The 
coordinates are given in units of the Einstein radius of the lens system. The
source trajectory is nearly perpendicular to the lens axis. There is a 
degenerate fit, with $\chi^2$ larger by $\Delta\chi^2 = 0.199$, that has a nearly
identical caustic configuration, except that the source crosses the outer cusp
instead of the inner one.
\label{fig-caustic_bin1}}
\end{figure}

The best fit model for MOA-bin-1, was found using the initial parameter
grid search method outlined in \citet{bennett-himag}, but a caustic centered
coordinate system was used, as described above. This procedure yielded
two nearly degenerate solutions, with a $\chi^2$ difference of 
$\Delta\chi^2 = 0.428$. These solutions have a nearly identical 
geometry, with the main difference being that the slightly favored
solution crosses the inner cusp of the planetary caustic, while the
second best solution crosses the outer cusp. The favored
solution is the one indicated in Fig.~\ref{fig-caustic_bin1}. There
are also disfavored solutions in which the source crosses the top or bottom
cusps  in Fig.~\ref{fig-caustic_bin1}. These differ from the inner and
outer cusp crossing solutions because the source must also
approach the primary lens caustic. If there are systematic errors in the baseline
part of the light curve, the $\chi^2$ might be improved if the
peak due to an approach to the primary caustic occurs at the
same time as a light curve bump due to systematic photometry
errors. When we are close to the Chang-Refsdal limit, the
main light curve features do not really constrain where such
a peak might occur, so the fitting code will naturally adjust the 
parameters to put this feature in the location that will minimize
the $\chi^2$. It is therefore important to correct systematic
photometry errors in the baseline photometry, as discussed
in Section~\ref{sec-syserr}. 

In this case, we find that the
solutions which cross the top or bottom cusp of the planetary
caustic, are disfavored by $\Delta\chi^2 = 17.5$. Formally, this
is enough to strongly exclude them. However, at $4135 < t < 4200$,
the top or bottom cusp crossing models are favored over 
the inner and outer cusp crossing models by $\Delta\chi^2 = 16.6$.
Most likely, this is simply due a small systematic error and/or
random noise, so the top or bottom cusp crossing solutions
should be disfavored by $\Delta\chi^2 \sim 34.1$. The parameters for these
solutions are substantially different from the parameters of the 
inner and outer cusp crossing models, with a mass ratio in
the brown dwarf regime, and a source that is fainter by a factor of
6-10. In addition, the non-planetary models imply an extremely faint
source in the OGLE $I$-band, and if this were correct, it would
imply a very blue source. We conclude that these models are not
consistent with the data, but we discuss them in more detail
in Appendix~\ref{sec-moa-bin1-nopl}.
However, we note that follow-up high angular resolution observations
can test this conclusion by determining the approximate brightness
of the source star. Models with a cusp crossing of a caustic with
different geometries are excluded by $\Delta\chi^2 >50$, and
they all have much fainter source stars.
We do not consider them further.

The parameters for the two remaining $\chi^2$ minima
are listed in Table~\ref{tab-mparam} in the center-of-mass coordinate
system. The parameters are the Einstein radius crossing time, $t_E$, the
planet-star separation, $s$, the angle between the source trajectory
and the lens axis, $\theta$, the planetary mass fraction, $\epsilon$,
the source radius crossing time, $t_\ast$, and
the time, $t_0$, and distance, $\umin$, of closest approach between the
source and center-of-mass. Both $s$ and $\umin$ are measured in units
of the Einstein radius. (The mass fraction is related to the mass
ratio by $\epsilon = q/(1-q)$.) The
caustic geometry of the best fit light curve model
(shown in Fig.~\ref{fig-lc_bin1}) is presented in Fig.~\ref{fig-caustic_bin1}.
The parameters for these two nearly degenerate models are quite 
similar, so the degeneracy will have little effect on the inferred physical
parameters of the planetary system. 

The similarity of the inner and outer cusp crossing models is an indication
that the lens model is nearly in the Chang-Refsdal limit 
\citep{chang-refsdal79,chang-refsdal84}, where it is only the 
gradient of the gravitational field of the host star that has
a significant influence on the light curve. In the Chang-Refsdal limit,
the planetary caustic, shown in Fig.~\ref{fig-caustic_bin1}, would be
reflection symmetric about a vertical axis.

The green, long-dashed curve in  Fig.~\ref{fig-lc_bin1} indicates the
light curve due to the host star alone, if there was no planet. The
peak magnification would only be 10\% without the planet. This
is not high enough for the isolated stellar lens to be detected because the
source is faint. The blue, short-dashed is the light curve due to the 
lensing effect of the planet alone (assuming
that the planet is located at the caustic centroid position). Without the
host star, the peak magnification would be $A_{\rm max} = 2.47$. This 
is a factor of 7.7 less than the observed peak magnification of 
$A_{\rm max} = 19.07$, but it would probably enough for an isolated
planet to be detected in the \citet{sumi11} analysis. However, Einstein
radius crossing time for the planet alone would be 
$t_{Ep} = \sqrt{\epsilon} t_E = 2.15\,$days. So, this event would not
add to the excess of events with $t_E < 2\,$days in the \citet{sumi11},
although events with $t_E \sim 2\,$days can be caused by lens
objects with either planetary or brown dwarf masses.

The physical properties of the MOA-bin-1LA,b planetary system are 
discussed in section~\ref{sec-prop}.

\subsection{MOA-bin-2}
\label{sec-moa-bin2}

\begin{figure}
\plotone{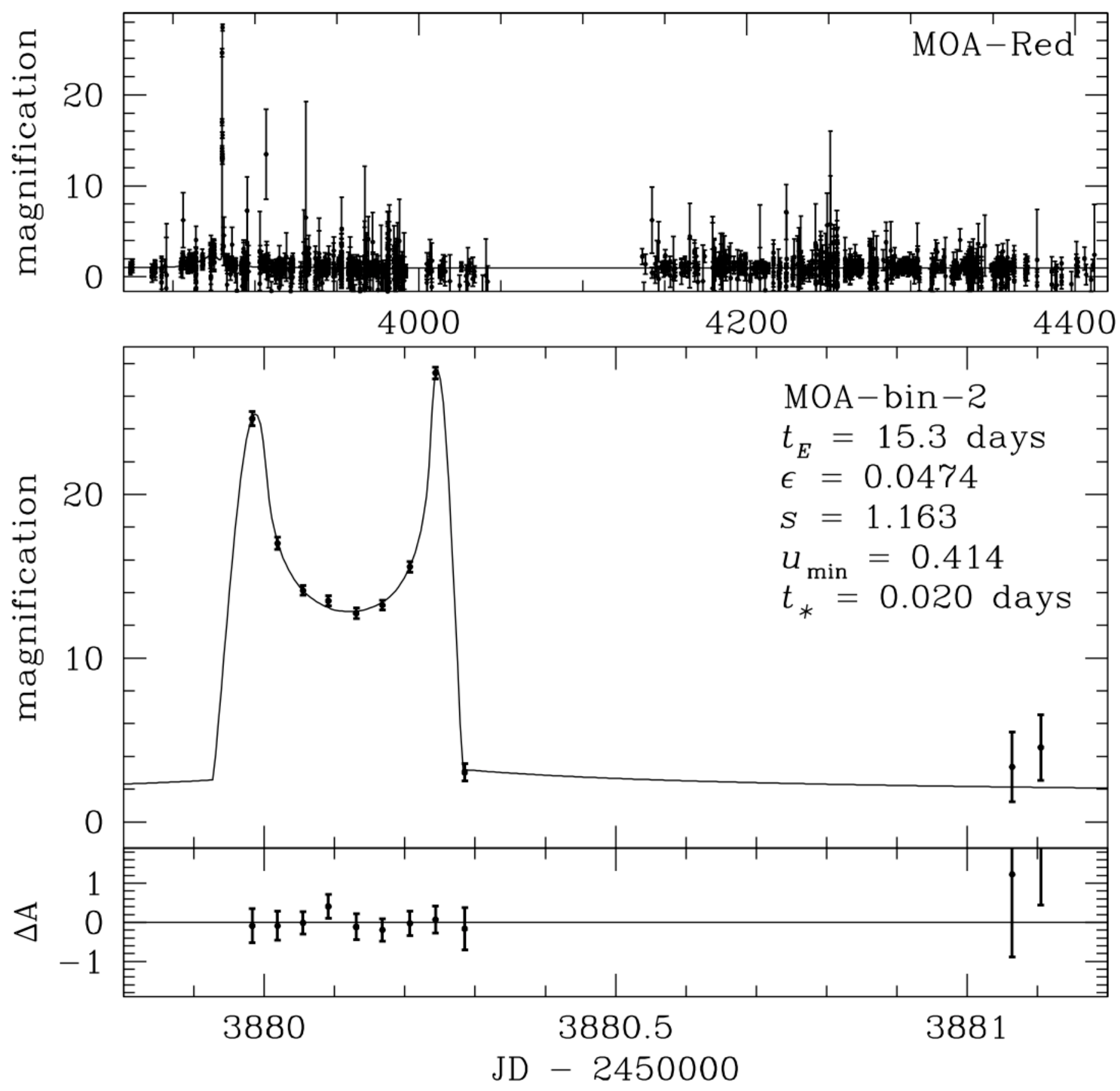}
\caption{The light curve of event MOA-bin-2 in field gb19 with a $\sim 50\,$min
sampling cadence. As in Fig.~\ref{fig-lc_bin1} top panel shows the 2006-2007
MOA-red data; the middle panel shows a close up
of the light curve peak, and the bottom panel shows the residuals
from the best fit. There are no OGLE data for this event.
\label{fig-lc_bin2}}
\end{figure}

Event MOA-bin-2 is located in field gb19, which 
had a $\sim 50\,$minute sampling cadence. The light
curve, shown in Fig.~\ref{fig-lc_bin2}, 
indicates a classic U-shaped caustic crossing
shape spanning $\sim 7\,$hours, and like the case of
MOA-bin-1, this is a light curve shape that is unique to microlensing.
However, there is no light curve coverage during the 4 days prior to
the caustic crossing. As a result, the model parameters are not 
very tightly constrained. The best fit model has a brown dwarf
mass ratio, $q = \epsilon/(1-\epsilon) = 0.0497$, and its other
parameters are listed in Table~\ref{tab-mparams}. However,
there is a
planetary mass ratio model with $q = 1.7\times 10^{-3}$ that is disfavored
by $\Delta\chi^2 = 4.08$, so a planetary secondary is not
strongly excluded.

\begin{figure}
\plotone{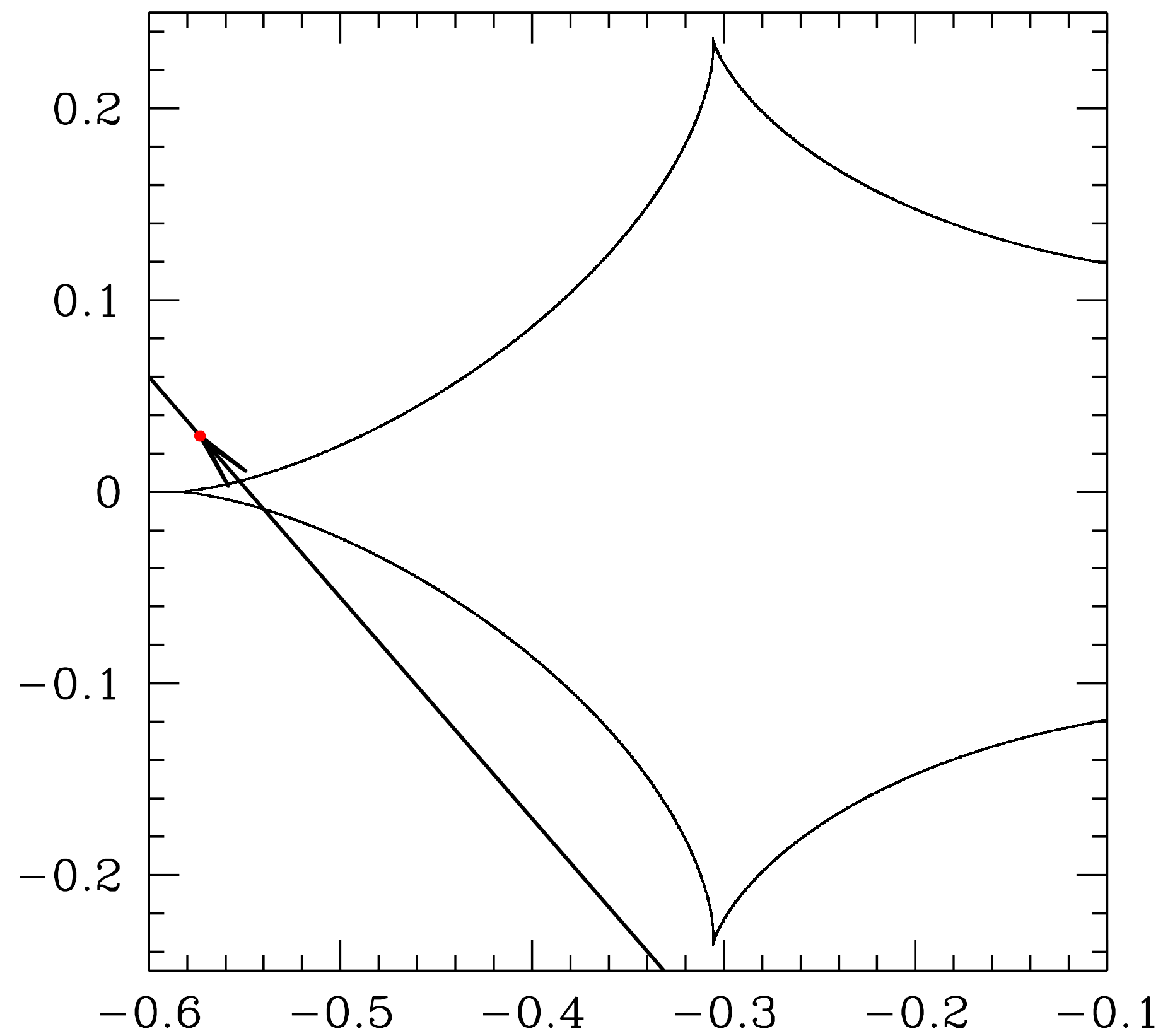}
\caption{The caustic geometry for the best fit MOA-bin-2 model. As in 
Fig.~\ref{fig-caustic_bin1}, the red circle indicates
the best fit source size, and the coordinates for this plot are centered at the 
center-of-mass of the full lens system.
\label{fig-caustic_bin2}}
\end{figure}

The caustic configuration for the best fit model is shown in 
Fig.~\ref{fig-caustic_bin2}. This model has a connected or
``resonant" caustic, while the planetary models cross the
same cusp (the most distant one) of a separated planetary
caustic.

\subsection{MOA-bin-3}
\label{sec-moa-bin3}

\begin{figure}
\plotone{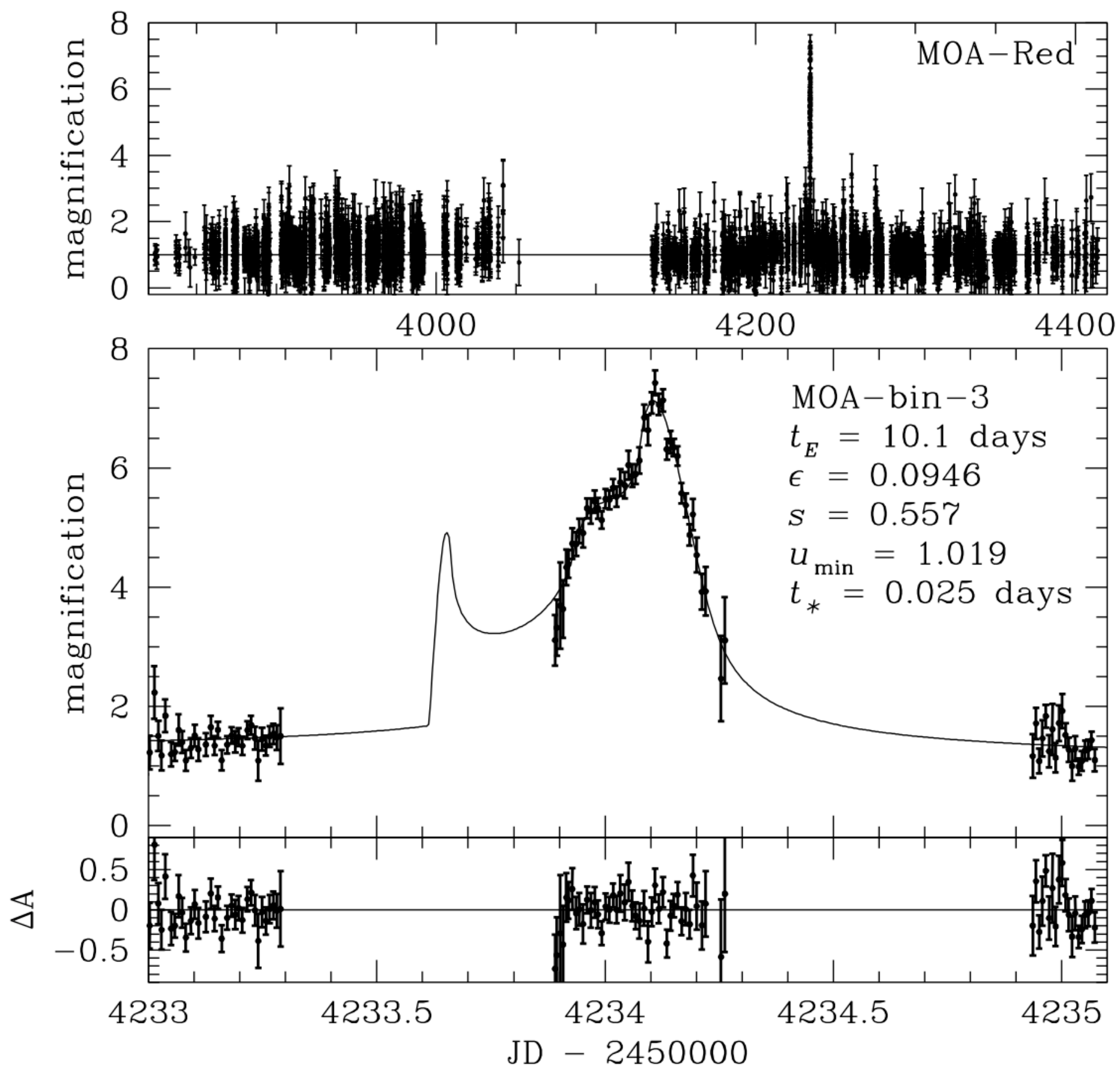}
\caption{The light curve of event MOA-bin-3 in field gb5 with a $\sim 10\,$min
sampling cadence. As in Fig.~\ref{fig-lc_bin1} top panel shows the 2006-2007
MOA-red-band data; the middle panel shows a close up
of the light curve peak, and the bottom panel shows the residuals
from the best fit. There are no OGLE data for this event.
\label{fig-lc_bin3}}
\end{figure}

Fig.~\ref{fig-lc_bin3} shows the light curve of MOA-bin-3,
which is the only binary event in our sample with a 
separation $s < 1$. A close separation binary is the one
configuration in which a system of two stars can give
rise to a very small caustic that is not surrounded by
a region of high magnification. So, as discussed by \citet{sumi11},
stellar binary lens systems with a separation $s \ll 1$, are the main 
microlensing background for the short duration single lens microlensing
events, which were used by \citet{sumi11} to infer a new 
population of isolated planetary mass lenses.

\begin{figure}
\plotone{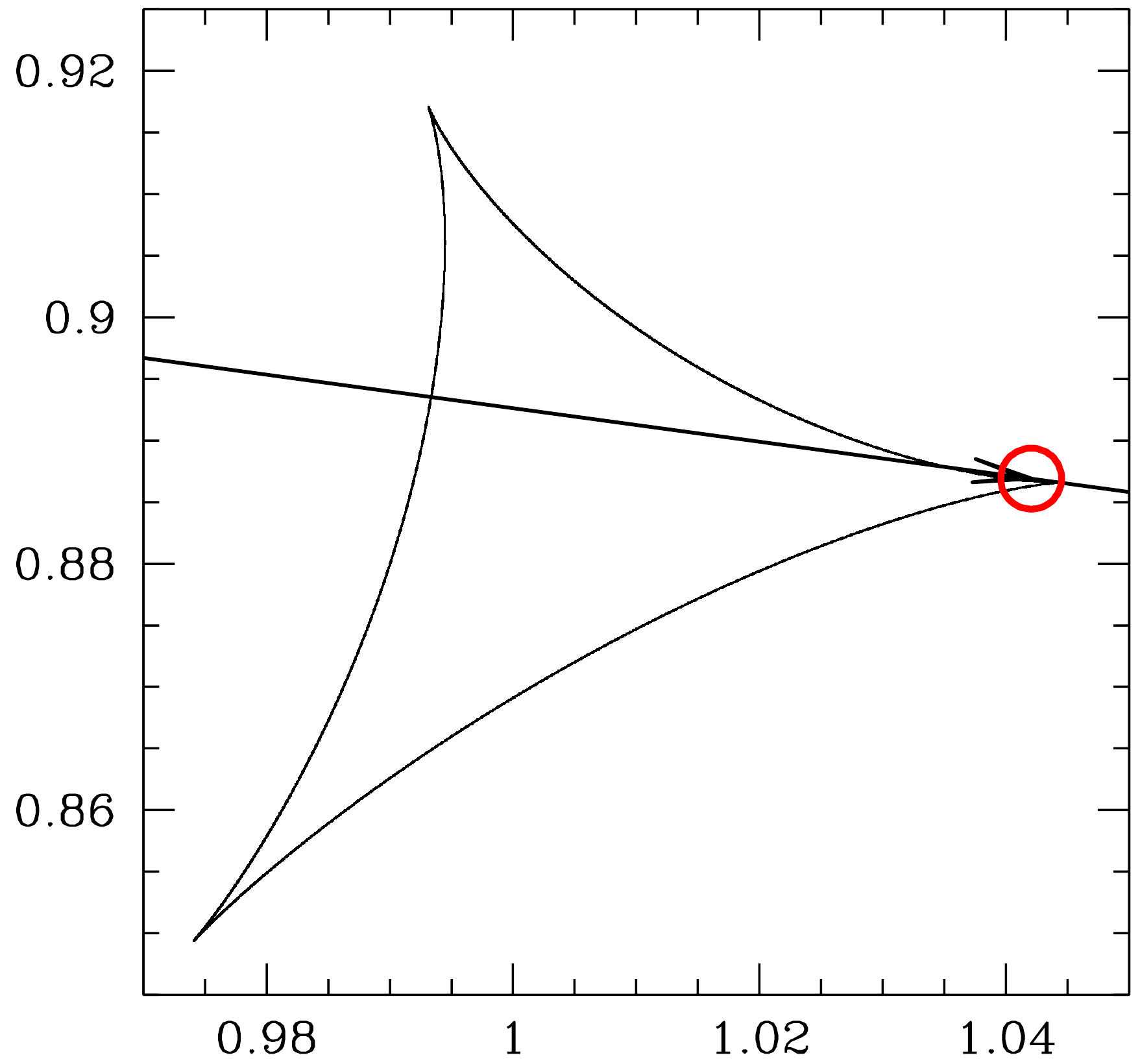}
\caption{he caustic geometry for the best fit MOA-bin-3 model. As in 
Fig.~\ref{fig-caustic_bin1}, the red circle indicates
the best fit source size, and the coordinates for this plot are centered at the 
center-of-mass of the full lens system.
\label{fig-caustic_bin3}}
\end{figure}

In the case of MOA-bin-3, the preferred fit has a moderate
separation of $s = 0.557$, which means that the caustic crossed
by the source (see Fig.~\ref{fig-caustic_bin3}) is not extremely
small, with a width (along the source trajectory) of $\sim 0.05\,R_E$.
But, the model compensates for this with duration of $\sim 10\,$days,
which is about half the typical event duration.

The mass ratio for this event is $q = \epsilon/(1-\epsilon) = 0.104$,
which implies that the secondary is a brown dwarf in the (likely)
case that the host star mass is $< 0.77\msun$. 

This event also has a wide separation model with $s = 1.609$ and 
$q = 7.7\times 10^{-3}$, but this model is disfavored by 
$\Delta\chi^2 = 19.7$. So, the close binary, brown dwarf secondary
model is probably the correct one.

\subsection{A Flare Star Fit by a Microlensing Model}
\label{sec-flare}

\begin{figure}
\plotone{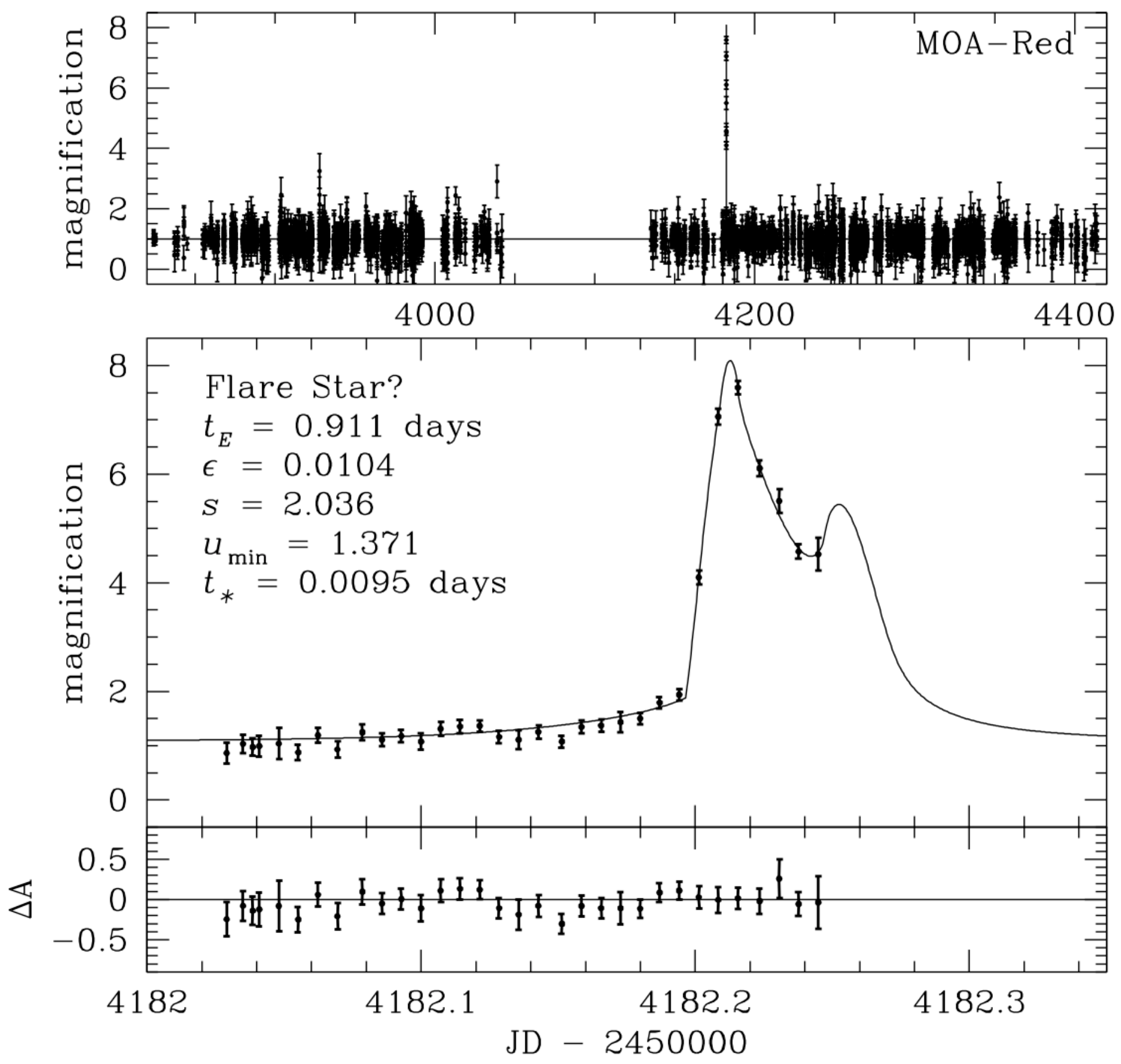}
\caption{The light curve of an event that is well modeled by an extremely short
binary lens light curve, with an Einstein radius crossing time of $t_E = 0.911\,$days.
The mass ratio is $q \approx 0.01$, which implies that if this model is correct, the
lens system would likely be an isolated gas giant planet with a moon that has a 
mass similar to Earth. However, the light curve features are not oversampled, and 
a foreground late M-dwarf is located at the position of the event. This implies that the
most likely interpretation is that this is a large amplitude M-dwarf flare 
\citep{kowalski10}.
\label{fig-lc_flare}}
\end{figure}

Fig.~\ref{fig-lc_flare} shows the light curve of the fourth event that is 
well fit by a very short binary lens light curve. If the microlensing model 
for this event was correct, it would be a remarkable discovery. The event
timescale is so short, $t_E = 0.911\,$days, that the primary lens is very likely to have
a planetary mass, similar to the mass of Jupiter. Thus, the secondary, 
which is nearly 
100 times smaller, would have a mass of a few Earth masses.

However, there are a number of reasons to be suspicious about this model.
It predicts a typical main sequence source brightness of $I\sim 19.5$, but
the source radius crossing time is quite small, $t_\ast = 0.0095\,$days.
If we assume a typical source radius of $\theta_\ast \sim 0.5\,$mas, then
we have a source-lens relative proper motion of 
$\mu_{\rm rel} \sim 19\,$mas/yr. This compares to a typical relative proper
motion of $\mu_{\rm rel} \sim 6\,$mas/yr, and it corresponds to a relative
velocity of $\sim 700\,$km/s if both the source and lens are in the bulge.
Most likely, it would indicate that lens is quite close to us, at a distance of
$\simlt 1\,$kpc.

A worrying feature of this small $t_\ast$ value is that it means that the
caustic crossing feature at $t \simeq 4182.2$ is not oversampled, 
despite the fact that this event is in high cadence field gb9, which has
a 10 minute sampling cadence. This 
means that we cannot claim to have confirmed a unique microlensing
feature for this event. In fact, there only 7 observations with fit
magnification $A > 2$, which compare to a total of 8 fit parameters:
5 binary lens fit parameters, 2 linear flux parameters (the source
and blend fluxes), and $t_\ast$. While there are also 5-10 additional
data points indicating a gentle rise just before the apparent 
caustic entry, the number of data points used to constrain the
model is not much larger than the number of model parameters. 
So, the fact that the data fit the model does not provide convincing
evidence that the model is correct. Furthermore, the caustic
exit feature is not observed, and caustic entrance features tend to
more closely resemble the light curves of eruptive variable stars, like
CVs and stellar flares.

The OGLE-III photometry for this field \citep{ogle3-phot} indicates that
the event is superimposed on a star with $I = 17.8$ and 
$V-I = 2.9$, with an apparent proper motion of $\sim 10\,$mas/yr.
In the microlensing interpretation, this could be the average of the combined proper 
motion of the the $I \simeq 19.5$ source star and the brighter foreground
star. The foreground star could not be the lens, but it could be a
stellar host associated with the planetary mass primary lens. It would
have to be $\simgt 10\,$AU from the lens to avoid detection.

Depending on the fraction of the total extinction to the bulge 
($E(V-I) \simeq 0.75$ \citep{pop_extinct}) that is in the 
foreground of this star, the observed color of this star 
would correspond to an M2-M4 dwarf. The observed brightening
would correspond to $\sim 1\,$mag for this foreground star (since it is
brighter than the ``host" star implied by the light curve model). This means
that a ``mega-flare" \citep{kowalski10} 
would be a possible explanation for  this observed light curve.
Studies of the frequencies of flares on ``inactive" M2-M4 dwarfs indicate
that a flare of this magnitude would be expected to occur about once
every $\sim 10\,$years \citep{kowalski09,hilton10,hilton11}. Thus, the
2-year MOA-II survey of $\sim 50\,{\rm deg}^2$, should see many
such mega-flares, and it is quite plausible that one such flare
could be fit by a microlensing model. Thus, this is our preferred
interpretation of this event. 

If the microlensing interpretation was correct and the foregound M-dwarf is
associated with the lens, there should be a
$I \sim 19.5$ source star moving away from the foreground 
M-dwarf at a rate of $\mu_{\rm rel} \simeq 19\,$mas/yr. Since the
microlensing event (if that is the correct interpretation) 
occurred in 2007, by 2012, the separation would be nearly $100\,$mas,
so high resolution observations with HST \citep{bennett06,bennett07}
or adaptive optics (AO)
\citep{sumi10,bennett-ogle109,janczak10,moa192_naco,m22ml}
should be able to detect the source moving 
away from the foreground star.

This event illustrates the importance of a high sampling cadence, so that
features that may be fit by a microlensing model are actually over-sampled.
The over-sampling allows for much a much stronger test of the microlensing
model as it will be much more difficult for the fitting code to adjust the 
parameters to match the light curve, unless the event is really due to
microlensing. We also note that stellar variability, like flares, are much
more likely to produce light curve that resemble caustic entrances
than caustic exits.

\section{The Physical Properties of MOA-bin-1LA, b}
\label{sec-prop}

Many of the light curve parameters presented in Table~\ref{tab-mparam} are
given in units of the Einstein radius, $R_E = \sqrt{4GM_LD_L(D_S-D_L)/(c^2D_S)}$,
where $M_L$ is the mass the lens system, and $D_L$ and $D_S$ are the masses
of the lens and source, respectively.
This means that the microlensing light curves can be modeled with a relatively
small number of parameters, but it also means that the most well measured 
parameters of a microlensing event are measured in units of $R_E$, instead
of physical units. In order to determine the physical parameters, it is
very helpful to be able to measure the angular Einstein radius, 
$\theta_E = R_E/D_L$. In most planetary events, finite source effects
are seen, so the source radius crossing time, $t_*$ can be measured.
Since we can estimate the angular radius of the source star, $\theta_*$, 
from its brightness and color, it is usually possible to determine
$\theta_E = \theta_* t_E/t_*$. For some planetary events 
\citep{bennett08,gaudi-ogle109,dong-ogle71,bennett-ogle109,muraki11},
it is possible to measure the projected Einstein radius, $\rep$, where the
projection is from the lens position to the Solar System. This is known
as the microlensing parallax effect \citep{gould-par1,macho-par1}, and
it is usually measured in long duration events, where the effect of the
orbital motion of the Earth can be measured in the light curve. When
this is measured, $M_L$ can be directly determined
\citep{gould-par1,bennett_rev,gaudi_rev},
\begin{equation}
M_L = {c^2\over 4G} \rep \theta_E \ .
\label{eq-m_rep}
\end{equation}

For the short duration events we present in this paper, there is no
data that can be used to measure the microlensing parallax effect,
so, we are left with the mass-distance relation,
\begin{equation}
M_L = {c^2\over 4G} \theta_E^2 {D_S D_L\over D_S - D_L} 
       = 0.9823\,\msun \left({\theta_E\over 1\,{\rm mas}}\right)^2\left({x\over 1-x}\right)
       \left({D_S\over 8\,{\rm kpc}}\right) \ ,
\label{eq-m_thetaE}
\end{equation}
(where $x=D_L/D_S$)
based on the measured value of $\theta_E$. In order to determine $\theta_E$,
we must first work out the color and magnitude of the MOA-bin-1S source star.
The magnitude in the non-standard MOA-red passband $R_M$ can be determined quite
precisely from the observed light curve. The MOA-red source
flux is determined from the best fit light curve,
shown in Fig.~\ref{fig-lc_bin1} (with parameters listed in Table~\ref{tab-mparam}),
and this can be compared to the standard Johnson $V$ and Cousins $I$ passbands
by comparison to the OGLE-III photometry catalog \citep{ogle3-phot}. This comparison
yields
\begin{equation}
R_M = I + 0.1485\left(V-I\right) \ ,
\label{eq-R_M}
\end{equation}
which differs somewhat from the transformation found by \citet{gould_col} 
for a comparison between the MOA-red passband and the OGLE-II photometry
catalog. In principle, we could use eq.~\ref{eq-R_M} as a 
color-color relation, $R_M - I = 0.1485(V-I)$ to relate the measured $R_M-I$
photometry to $V-I$ as advocated by \citet{gould_col}.
(In Appendix~\ref{sec-src_col}, we apply this method to five of the ten isolated
planet events from \citet{sumi11}.)
As discussed in Section~\ref{sec-moa-bin1}, for MOA-bin-1,
there are no OGLE data at
relatively high magnification, and they are not used in the fit due to concerns
about blending-related systematic errors. 
Therefore, we estimate the source
star color following the procedure of \citet{bennett08}.

Fig.~\ref{fig-cmd_bin1} shows the OGLE-III CMD  
\citep{ogle3-phot} of the stars within 1 arc minute of the MOA-bin-1S source star.
The centroid of the red giant clump is located at 
$(V-I)_{\rm rc} = 3.77$, $I_{\rm rc} = 17.51$,
which compares to the expected absolute magnitudes \citep{bennett-ogle109} of 
$M_{I\rm rc} = -0.25$ and $M_{V\rm rc} - M_{I\rm rc} = 1.04$. Since the event
is located at a Galactic longitude of $l = -0.111^\circ$, we assume a distance of
$D_S = 8.0\,$kpc or a distance modulus of $DM = 14.52$. This implies
extinction of $A_I = 3.24$, $A_V = 5.97$, and $R_{VI} = A_V/(A_V-A_I) = 2.19$,
since $A_I = I_{\rm rc} - M_{I\rm rc} -DM$ and
$A_V = V_{\rm rc} - M_{V\rm rc} -DM$.
This $R_{VI}$ value is quite typical of Galactic bulge fields
\citep{udalski-ext}, but the total extinction is much higher than is typical
for the fields of observed microlensing events. The high extinction is also
apparent from direct inspection of Fig.~\ref{fig-cmd_bin1}, as the OGLE
catalog becomes highly incomplete (in the $V$-band)
at less than 1 mag fainter than the red clump centroid.

We can estimate the source radius, following 
\citet{bennett08}, by transforming the Baade's Window 
CMD observed with HST by \citet{holtzman98} to the extinction and
distance appropriate for the field of MOA-bin-1. When this is done, we find
that there is considerably more scatter in the MOA-bin-1 CMD 
than in the HST Baade's WIndow CMD. This is likely to be due to
excess differential reddening in the MOA-bin-1 field. Therefore, we add a dispersion
of 0.101 mag along the reddening line, and the resulting HST CMD is represented
by the points shown in green in Fig.~\ref{fig-cmd_bin1}.

The best fit magnitude for the MOA-bin-1S source star is $R_{Ms} = 20.73$ 
(see Table~\ref{tab-mparam}). As shown in Fig.~\ref{fig-cmd_bin1}, this is
less than three $I$ mag fainter than the centroid of the red clump.
This implies that the source could be a relatively young bulge
main sequence star \citep{bensby11}, a turn-off star, or a 
sub-giant near the base of the giant branch. 
Comparison with the reddened 
HST CMD gives an $I$ magnitude of $I_s = 20.23$ and a
color of $(V-I)_s = 3.38 \pm 0.15$ for this particular
model. The corresponding dereddened values are therefore
$I_{s0} = 16.99$ and $(V-I)_{s0} = 0.65 \pm 0.16$.
The surface brightness-color relation of \citet{kervella08} yields
a source radius of $\theta_* = 1.23 \pm 0.18\,\mu$as, where the
error bar only reflects the effect of the uncertainty in the source color.
There is an additional uncertainty in the source radius due to the 
uncertainty in the source star brightness, but this is strongly correlated
with the other event parameters. We account for this uncertainty separately
in our Markov Chain Monte Carlo (MCMC) calculations, by adjusting
the value of $\theta_*$ to account for the variations in the source
brightness for each different set of model parameters.
As indicated in Table~\ref{tab-mparam}, the outer cusp model
has a best fit source magnitude of $R_{Ms} = 20.75$, so these
models predict a slightly smaller source,
$\theta_* = 1.21\,\mu$as. However, this solution also has 
shorter source radius crossing time, so the 
lens-source relative proper motion values for this solution is 
slightly larger, at $\mu_{\rm rel} = \theta_*/t_* = 8.9\pm 1.4$ for 
the outer cusp solution than the value, $\mu_{\rm rel} = 8.8\pm 1.4$
for the best fit model. These values are a little higher than the 
$\sim 6\,$mas/yr typical of bulge lensing events, but these $\mu_{\rm rel}$
are consistent with a lens system either in the disk or the bulge.

The implied angular 
Einstein radius values are $\theta_E = \theta_* t_E/t_* = 0.77 \pm 0.11\,$mas
for the best fit (inner cusp crossing) model and
$\theta_E = 0.76\pm 0.11\,$mas for the best outer cusp crossing
model. The effect of the larger $t_E$ and smaller $t_*$ compensate
for the smaller $\theta_*$ for this model. These $\theta_E$ values
can be used to constrain the mass of the lens system using
eq.~\ref{eq-m_thetaE}. 

\citet{macho-95b30} present a method to use
eq.~\ref{eq-m_thetaE} to make a Bayesian
estimate the probability distribution of the
distance, $D_L$, and mass, $M_L$, based on a Galactic model of the 
stellar kinematic and density distribution. We have supplemented
this method by allowing the source distance, $D_S$, to vary, and
used the Galactic model parameters from \citet{bennett02}.
This yields a probability table 
as a function of relative proper motion, $\mu_{\rm rel}$, source distance,
$D_S$, and the lens system distance, $D_L$. This probability table
is used as an input to a program that sums the MCMC 
likelihood results in order to properly weight the different light curve
models with the probability of the parameters that they imply.
This routine 
determines the probability distribution of the physical parameters,
which depend on both the fit parameters and the unmeasured
parameters $D_L$ and $D_S$. For the relative proper motion,
$\mu_{\rm rel}$, which is measured, we use the galactic model weights
to marginalize over the measurement uncertainty. For each model
in the MCMC, we have a value of $t_E$, which can be combined
with the relative proper motion to yield the angular
Einstein radius, $\theta_E = \mu_{\rm rel}t_E$. This
allows us to determine the lens mass using eq.~\ref{eq-m_thetaE},
since we are marginalizing over $D_L$ and $D_S$. Once the lens
mass is determined, we can apply a mass prior, using the 
power-law mass function of
\citet{sumi11}. (This is mass function \# 3 presented in the 
Supplementary Information.) The advantage of this mass
function is that it includes possible brown dwarf and even planetary
mass hosts, which could plausibly host a planetary mass ratio
companion.

The results of these MCMC calculations are summarized in
Table~\ref{tab-pparam}, which gives the median values and
$1\,\sigma$ and $2\,\sigma$ confidence level limits on the 
planetary system distance, $D_L$, host star mass, $M_*$, 
planet mass, $m_p$, and the semi-major axis, $a$. Both
the OGLE and MOA data indicate that the MOA-bin-1 source star
is blended with a star with an $I$ magnitude of $I_b = 19.66$,
which can be used as an upper limit on the source brightness.
Due to the high dust extinction in this field, however, this
constraint does not exclude host stars less massive
than $1\,\msun$. The bulge turn-off mass is
$\sim \msun$, but there are some young stars in the 
bulge \citep{bensby11}. In order to avoid excluding these
possible host stars with an {\it a priori} assumption,
we have modified the \citet{sumi11} mass function \# 3
by adding a population of such stars with a mass
function that scales like $dN/dM \propto M^{-4}$ for 
$M > 1\,\msun$.

Table~\ref{tab-pparam} indicates the host star is most likely
to be a an late K-dwarf, but the $2\,\sigma$ range spans virtually
the entire main sequence available in the bulge. The planet
is quite massive, at $m_p = 3.7\pm 2.1\,M_{\rm Jup}$, probably
the most massive planet yet found by microlensing. It may also
be the widest separation planet found by microlensing, although
one of the two degenerate solutions for MOA-2007-BLG-400Lb 
\citep{dong-moa400} does have a wider separation.

Previous microlensing work has shown that sub-Jupiter-mass
gas giants are rather common in orbit around M-dwarfs \citep{gould10},
and has also found several cases of super-Jupiter-mass planets
orbiting M-dwarfs \cite{udalski05,dong-ogle71,batista11}. 
In contrast, radial velocity studies \citep{johnson07,bonfils11} have
found a low occurrence rate of super-Jupiter-mass planets orbiting
M-dwarfs. However, most of the radial velocity sensitivity is to
planets in shorter period orbits than the microlensing discoveries, so
it is not clear that there is a discrepancy between these results.
The core accretion theory does seem to predict that only a small fraction
of M-dwarfs will have massive gas-giant planets
\citep{laughlin04,ida05,kennedy_snowline,dangelo_book}, so
If the MOA-bin-1Lb planet and its host have the median predicted masses,
then this might suggest that this planet formed by gravitational 
instability \citep{boss06} instead of core accretion. However, the
error bars on the MOA-bin-1L host star mass are quite large, so 
a solar-type host is also possible.

The uncertainty in the host star properties can be resolved with
HST follow-up observations, which should be able to detect the
host star if it has a mass of $\simgt 0.4\msun$ using the 
color dependent centroid shift \citep{bennett06,dong-ogle71} or
the image elongation \citep{bennett07} methods. These methods
should be able to detect the host star at
current separation of $\sim 50\,$mas assuming a mass of $\simgt 0.4\msun$.
It may also be possible to detect the host star with AO
infrared imaging \citep{bennett-ogle109,sumi10,janczak10,moa192_naco}, since
a diffraction limited image could resolve the source and planet host
stars. However, since we have no infrared imaging while the event was in
progress, there is some potential that IR images may not be able to 
distinguish the source and planet host stars. But, assuming that we
do not have this problem, the follow-up observations would also be able
to provide a measurement of the source star color, which would improve
our estimate of $\theta_*$ and allow the determination of the 
host star mass using eq.~\ref{eq-m_thetaE} and a mass-luminosity
relation as discussed in \citet{bennett06,bennett07}.

\section{Implications for Planets in Wide Orbits}
\label{sec-wide}

The bona fide planetary event, MOA-bin-1, is an example of a previously
unseen type of planetary microlensing event - an event nearly in the
Chang-Refsdal limit \citep{chang-refsdal79,chang-refsdal84}, 
where the main effect of the host star is to create
the planetary caustic. 
In order to establish the precise statistical implications of the discovery
such events,
we should calculate their detection efficiency. We haven't
done this, but a detailed consideration of the MOA-bin-1 allows us
to make a rough estimate of the detection efficiency for such events.
The green dashed curve in Figure~\ref{fig-lc_bin1}
indicates that the magnification due to the host star alone is so small,
$A_{\rm max} = 1.10$, that the event would be undetectable,
without the planet. (The source
is blended with another star that is a magnitude brighter, so the apparent
peak magnification is only $A_{\rm max} \sim 1.03$.) If the planet was
isolated, without a host star, it would generate the light curve
indicated by the blue-dashed curve in Figure~\ref{fig-lc_bin1}, which
has a peak magnification of $A_{\rm max} = 2.58$. This is high enough
so that the event would have still been detected in the short timescale
analysis of \citet{sumi11}, but it is much smaller than the observed
peak magnification of $A_{\rm max} = 19.1$.

This illustrates why the detection efficiency for 
Chang-Refsdal-like events is likely to be higher than for single lens
events with the equivalent effective parameters. For MOA-bin-1,
the observed light curve is much brighter than the equivalent single
lens light curve. This is a generic situation because
the caustics of Chang-Refsdal-like lens systems spread the region
of relatively high magnification ($A\simgt 10$) over larger area
than for the equivalent single lens events. The area of extremely
high magnification $A\simgt 100$ is reduced, but this is likely to
affect the detection efficiency less than the larger area at $A\simgt 10$.
One might also worry that our subjective procedure to select binary
events from the low-level event candidates of the \citet{sumi11} 
sample could have missed some events, but we feel that this
is unlikely since we have modeled all the events that remotely
resemble short duration binary lens events.

With the assumption that the detection efficiency for 
Chang-Refsdal-like events is higher than for single lens events, we
can combine the results of this paper with those of \citet{sumi11} to
determine properties of the semi-major axis distribution of this 
newly discovered planet population, assuming that they are in
very distant orbits about their host stars. We have found only one
Chang-Refsdal-like event that is a bona fide microlensing event due
to a planetary mass lens, MOA-bin-1. However, the effective timescale of this
event is actually too large to qualify for the \citet{sumi11} sample. If we take away
the planetary host star for this event, then the planet-only Einstein
radius crossing time becomes $t_{Ep} = \sqrt{\epsilon}t_E = 2.20\,$days, which
is above the \citet{sumi11} cut of $t_E < 2\,$days. Thus, the selection
criteria which yield 10 isolated planetary mass microlenses find no planets
at very wide separations with $s \geq 2$. (Note that MOA-2007-BLG-400 
\citep{dong-moa400} is
an event in this sample, and it has a planet which may be in a very wide or
a very close orbit, but the planetary signal was not seen in the MOA data
and would not have been seen without the high magnification due to
its host star.) 

We can now use this result in combination with the lower limits on the 
separations of possible host stars for the 10 short single lens events
of \citet{sumi11} to determine the properties of the planet separation
distribution, under the assumption that these 10 isolated planetary
mass objects are actually bound in distant orbits around stars.
The lower limits on the separations of the 10 events with $t_E < 2\,$days
are listed in Table 1 of \citet{sumi11},
and they are $s > 7.0$, 3.3, 3.6, 3.1, 2.4, 4.8, 5.2, 4.8, 3.4, and 15.0
for events MOA-ip-1 through 10, respectively. These are
measured in units of the Einstein radius of the possible host star.
We can now use these lower limits to compare to possible distributions
of wide separation planets to the data. We chose a very simple model,
\begin{equation}
{dN\over ds} \propto s^n \ \ \ {\rm for} \ \ s_0 < s < s_1 \ ,
\label{eq-dNds}
\end{equation}
where the number of planets as a function of separation scales as
a power law between inner and outer limits, $s_0$ and $s_1$. To
conform with the limits from \citet{sumi11}, these limits are all taken
to be in units of the Einstein radius, $R_E$, of the possible
host star. We select a lower separation limit of $s_0 = 2$, as this
guarantees that star itself will not have a strong microlensing
effect, so that the detection efficiency calculations of \citet{sumi11}
will be applicable instead of the very different calculations needed
to determine the detection efficiency to find planetary signals in
events where the magnification of the host star dominates
\citep{gould10,cassan12}.

Perhaps, the most reasonable guess for $n$ in equation~\ref{eq-dNds}
would be $n = -1$, so that there is an equal probability of 
finding a planet in a logarithmic interval (between $s_0$ and $s_1$).
This is the integer value of $n$ that is most compatible with observations,
and it is the only integer value consistent with planets at a wide
range of separations. Also, since the binding energy of the planet to its
host star scales as $s^{-1}$, this is the choice that allows the the
highest density of planets in stable orbits.
With the selections, $n = -1$ and $s_0 = 2$, 
we now only need to specify $s_1$ in order to work out the
probability that some number of the 10 (assumed bound) planets
will have their host stars detected in the \citet{sumi11} microlensing
data. With these values selected, it is then trivial to simulate
$10^6$ simulations of the \citet{sumi11} 10-event sample and
determine what fraction of these simulated observations reproduce
the observed number of 0 out of 10 detectable host stars.
Since, we seek to set a limit on the planetary distribution to be
consistent with the observations,
we select $s_1$ values that will yield a probability of 5\% or
10\% to detect 0/10 host stars. As shown in
Table~\ref{tab-hostprob}, $s_1 = 60$ yields a probability of
$P(0/10) \simeq 5\,$\%, and $s_1 = 135$ yields a probability
$P(0/10) \simeq 10\,$\%. So, these planet separation distributions
represent the 95\% and 90\% confidence levels with our
assumptions on the form of the planet separation distribution and
that all planets are bound.

One possible objection to this choice of $n = -1$ is that the
abundance of isolated planets found by \citet{sumi11},
$1.8{+1.7\atop -0.8}$ per main sequence star, is somewhat larger
than expected. So, perhaps, we should consider a distribution with
more outer planets than we would expect. Another choice that would
put more planets in outer orbits is $n = 0$, a uniform distribution
is separation instead of in the logarithm of the separation. This choice
is shown in the last two columns of Table~\ref{tab-hostprob}. Since these
distributions are more heavily weighted toward outer planets, the 
$s_1$ values corresponding to the 95\% and 90\% confidence level
limits are much smaller, at $s_1 = 18$ and 21, respectively.

While the outer cutoffs for these $n = -1$ and $n = 0$ distributions are
quite different at the 95\% or 90\% confidence levels, the median values
of these distributions (equation~\ref{eq-dNds}) are much closer. For
95\% confidence, at $n = -1$, we have $s_1 = 60$, which implies
a median of $s_m = 11.0$. At $n=0$, the 95\% confidence value for
$s_1 = 18$, which gives a median of $s_m = 10.0$. At 90\% confidence,
the $s_1$ values are 135 and 21 for the $n=-1$ and $n=0$ models, respectively,
and these give median values of $s_m = 16.4$ for $n=-1$ and $s_m = 12.5$
for $n=0$.

Now, for a typical lens star, a two-dimensional separation of
an Einstein radius ($s=1$) corresponds to an orbital semi-major axis
of $\sim 3\,$AU. Thus, the 95\% confidence limit of $s_m > 10.0$ from the 
$n=0$ model implies that a distribution
of bound planets that could explain the \citet{sumi11} results 
if it had a median semi-major axis of $a_m > 30\,$AU. With the
$n=-1$ model, the limit would be slightly stronger, $a_m > 33\,$AU.
While we have only considered a couple of simple models for the 
separation distribution of planets in wide orbits, the models that
we do consider do bound the power-law models that best match
the observed distributions of exoplanets \citep{cumming08,cassan12}.
Thus, it seems likely that a substantial fraction of the \citet{sumi11}
population of isolated planets must be unbound or in orbits beyond
$ 30\,$AU.

\section{Summary and Conclusions}
\label{sec-conclude}

In this paper, we have presented the analysis of 4 events from the
MOA Project's 2006-2007 short event analysis, which were well-fit by
binary lens light curve models. Two of these events, have best fit
binary microlensing models, in which the signal is dominated by
a planetary mass secondary with a distant primary that affects the
light curve primarily through the gradient of its gravitational field,
but we do not consider them both to be bona fide microlensing events.
One of these events, MOA-bin-1, has a an oversampled, characteristic cusp-crossing
feature, which makes it obvious that the microlensing interpretation is correct. The
other event with a best fit planetary mass microlensing light curve has an
apparent caustic crossing feature that is much more rapid than for typical 
microlensing events, and as a result, the feature is only marginally sampled
by the high cadence MOA images. The number of observations that are
much brighter than the baseline brightness is not much higher than the
number of model parameters needed to describe them, so the fact that the
light curve is well fit by a microlensing model does not imply that the model
is likely to be correct. Furthermore, it is located at the position 
of a foreground  M-dwarf that is expected to have a ``mega-flare" of
a similar magnitude about once every ten years. This is much higher than
the expected lensing rate, so we have concluded that this event is 
unlikely to be a microlensing event. This event illustrates the 
importance of oversampling the microlensing features in the 
light curve. Oversampling implies that there are many more
data points on the light curve features than there are model 
parameters that can be adjusted to fit the data. This allows the
additional data points to be used to establish the microlensing
nature of the event, as occurs for MOA-bin-1 (and to some extent
for MOA-bin-2).

We have also done a preliminary investigation of the implications
of the search for Chang-Refsdal-like planetary events, using a
rather crude estimate of the detection efficiency for our somewhat
subjective event selection procedure. In particular, we have assumed
that the detection efficiency for such events is not smaller than the
detection efficiency for the equivalent short single lens events. 
The arguments given in Section~\ref{sec-wide} indicate that this
is likely to be the case, but this has not been directly demonstrated.
If all of the isolated planetary mass microlensing events found
by \citet{sumi11} were due to bound planets with a simple 
power-law distribution, we find that the median star-planet
separation must be $\simgt 30\,$AU. 

We plan a more detailed analysis of this question in a future paper
with three times larger sample of events, and with a detailed analysis
of the detection efficiencies. We expect that this analysis will
provide a significantly stronger limit than we have presented here
for a couple reasons. First, here we
have assumed that the detection efficiency for the wide separation planets
is the same as for isolated planets, but in fact, for events like MOA-bin-1, the
detection efficiency for the wide separation planets is likely to be higher.
So, the failure to detect any wide orbit planets with $t_{Ep} < 2\,$days is
likely to be more significant than we have assumed.
Second, a much larger sample of short duration microlensing events with
$t_E < 2\,$days has been obtained in the years since 2007 with the MOA-II
alert system \citep{bond01} and the OGLE-IV Early Warning System \citep{ogle-ews},
but no other examples of wide separation planets, beyond MOA-bin-1 
(with $t_{Ep} = 2.2\,$days) have been
seen. This suggests a more rigorous future analysis
using calculated wide-planet detection efficiencies and the
full sample of 2006-2012 MOA-II and 2010-2012 OGLE-IV data should
a much more stringent lower limit on the median semi-major axis of
these isolated planetary mass lenses, unless other wide separation
planet events have been missed.

Finally, an alternative method to probe the possibility of host
stars for the isolated planets of \citet{sumi11} is with follow-up
HST imaging of the current sample of $\sim 30$ such events, 
plus a similar size sample of events with $2\,{\rm days} < t_E < 4\,$days.
While the events with $t_E < 2\,$days are thought to be mostly
due to planets, most of the events with $t_E \sim 3$ or 4 days
are likely to be due to brown dwarfs. If a substantial fraction
of the isolated planet population has host stars, then 50\% or
more of the host stars should be detectable in HST images, and
multiple epoch HST imaging can confirm that the candidate
host stars have proper motions consistent with our
expectations for host stars using the methods of
\citet{bennett06,bennett07}. On the other hand, we would
expect that almost all of the brown dwarf lenses should not
have host stars, due to the brown dwarf desert. So, HST observations
of $t_E \sim 3$ and 4 day events should not reveal host stars.
(A few events may be due to high proper motion stars, but
these are readily identified with multiple HST frames.)
The combination of HST observations with the analysis of
the post-2007 microlensing data should teach us a great
deal about the nature of this newly discovered isolated
planet population.

\acknowledgments 
We thank the OGLE collaboration for permission to use their
unpublished data for the events presented in this paper.
We acknowledge the following support:
NASA-NNX10AI81G and NSF AST-1009621 (DPB);
JSPS18253002 and JSPS20340052 (FA); 
JSPS 19340058 (YM) and MEXT 14002006 (YM).

\appendix
\section{Non-planetary Models for MOA-bin-1}
\label{sec-moa-bin1-nopl}

\begin{figure}
\plotone{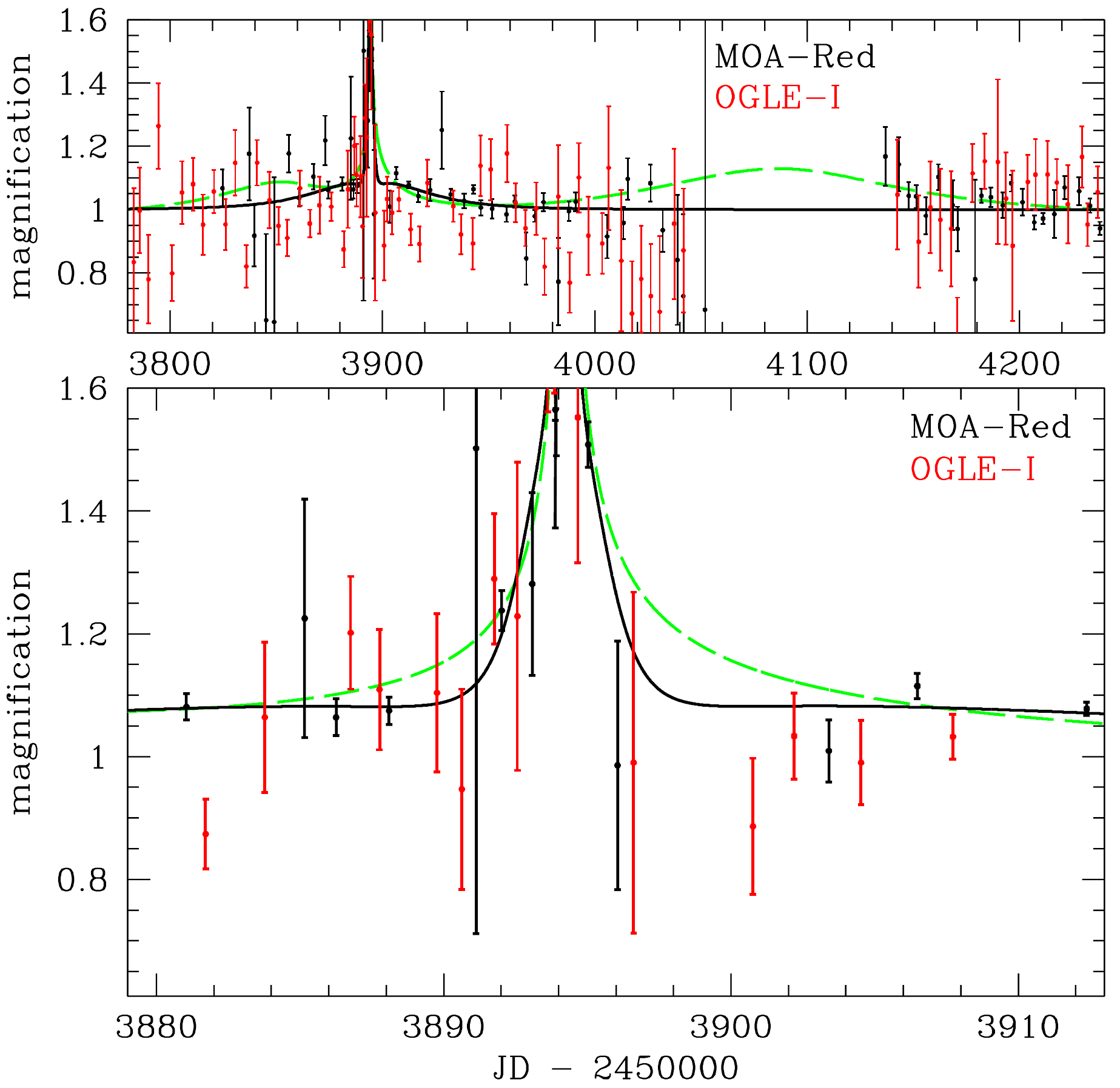}
\caption{Comparison of planetary
(black) and non-planetary (green dashed) light curve models for
event MOA-bin-1. The MOA and OGLE data are binned with 
different bin sizes depending on the time difference from
the cusp crossing peak. There is no binning within 12 hours from
the peak. Between 0.5 and 4.5 days, the data are binned in
1-day bins, and between 4.5 and 10.5 days, 2-day bins are used.
The remaining data is binned in in 5-day bins.
\label{fig-lc_bin1nopl}}
\end{figure}

As discussed in Section~\ref{sec-moa-bin1}, there are are non-planetary
models for MOA-bin-1 that can describe the cusp crossing feature but 
are excluded by the observations of the light curve outside of the cusp
crossing peak. These solutions have a similar caustic structure to the
best solution, indicated in Fig.~\ref{fig-caustic_bin1}, but the source 
crosses the top or bottom cusp instead of the inner or outer one. 
Because these solutions involve source motion that is nearly parallel
to the lens axis, they are likely to have additional peaks due to a
close passage to the cusp associated with the other mass (in this case,
the primary mass). However, because of the near Chang-Refsdal nature
of the light curve, there are only weak constraints on the location and
strength of such a peak in the light curve. This allows such a feature
to be attracted to systematic error features that might exist in the
light curve. As mentioned in Section~\ref{sec-moa-bin1} and shown
in Figure~\ref{fig-lc_bin1nopl}, the non-planetary solution for MOA-bin-1
gets a $\Delta\chi^2 = 16.3$ improvement from data in late-February, early-March, 2007
($t\sim 4165$).

We believe that this feature at $t\sim 4165$ is likely to be due to a systematic
photometry error. As discussed in Section~\ref{sec-syserr}, the MOA photometry for
MOA-bin-1 shows evidence of a systematic error related to the direction
and amplitude of differential refraction. Our empirical correction of this effect
has reduced the significance of this apparent feature from $\Delta\chi^2 \sim 95$
to  $\Delta\chi^2 = 16.3$. It seems likely that the remaining improvement
could be due to an imperfect correction. 

One obvious way to check the possibility that this feature is due to a systematic
error in the MOA data is to compare to the OGLE data. However, as mentioned
in Section~\ref{sec-moa-bin1}, there is 
also a systematic error that affects the OGLE data. OGLE uses an implementation
of difference imaging \citep{ogle-pipeline} with photometry done at the centroid
of stars identified in the reference frame. But, because the central Galactic bulge
fields are very crowded, the faint stars identified in the OGLE frames are typically
blends of several stars, so the centroids of the identified stars do not line
up with the actual position of the faint source stars. Often, the photometry can
be significantly improved by identifying source star position from the centroid
in a good seeing difference image, taken when the source is highly magnified. 
Unfortunately, the source is not seen at very high magnification in the OGLE 
data, and the three OGLE data points at moderate magnification of $A \sim 1.7$
where taken in poor seeing. So, identification of the optimal centroid of the 
source in the OGLE data is difficult, and this is probably why
optimum centroid photometry of the OGLE data is not available.

Without the optimum centroid photometry, there are two choices for dealing 
with the OGLE photometry. The simplest choice is to to simply fit the OGLE
data as an independent data set. This gives the results that were reported in
Section~\ref{sec-moa-bin1}. However, we can use the methods described in
Appendix~\ref{sec-src_col} to work out the source color implied by these
models, and these indicate unreasonably blue source colors. For the
best fit model, with a crossing of the outer cusp of a planetary mass
secondary, this gives $(V-I)_s = -0.1\pm 1.2$, while for the best fit
upper or lower cusp crossing model (with a non-planetary mass ratio)
the implied color is $(V-I)_s = -4.1\pm 1.4$. As shown in 
Figure~\ref{fig-cmd_bin1} and discussed in Section~\ref{sec-prop}, the
real source color is $(V-I)_s = 3.38 \pm 0.15$, which is 2.9-$\sigma$
redder than the best fit (planetary) model and 5.3-$\sigma$ redder than
the best upper or lower cusp crossing (non-planetary) model. These
unreasonable colors are an indication of a problem with the OGLE 
photometry centroid.

If we require that the source brightness in the OGLE $I$-band be consistent 
with a normal source star (\ie\ with $(V-I)_s = 3.38 \pm 0.15$), we find that
$\chi^2$ is increased by $\Delta\chi^2 = 7.9$ for 2373 OGLE $I$-band
observations. This is another indication that the systematic error driving
the source color toward an unreasonably blue value is just under 
a 3-$\sigma$ effect.

Since we know that a normal source star must have a color of 
$(V-I)_s = 3.38 \pm 0.15$, it is sensible to enforce such a constraint
on the different models for MOA-bin-1. Because the non-planetary,
upper and lower cusp crossing models predict a source color that
is too blue by more than 5-$\sigma$, we should expect that these
non-planetary models should be disfavored by the requirement that 
the lensed source have a reasonable stellar color, and this is
indeed the case. The $\chi^2$ difference between the best fit model
and the best fit upper or lower cusp crossing model increases
from $\Delta\chi^2 = 17.5$ to $\Delta\chi^2 = 54.4$. The lower
magnification part of these two models are shown in Figure~\ref{fig-lc_bin1nopl},
with the best fit model shown in black and the best upper or lower
cusp crossing model shown as a dashed green curve. The data shown
in this Figure is binned for clarity, but all the fits are done with the
original unbinned data. 

The parameters for the best fit planetary model with this constraint
are somewhat different from the values listed in Table~\ref{tab-mparam}.
With this constraint, the best fit has $t_E = 22.5\,$days, $t_0 = 3894.4$,
$u_{\rm min} = 1.659$, $\theta = 1.565$, $\epsilon = 0.0078$,
$t_\ast = 0.0489\,$days, and $R_M = 20.65$. The best fit upper or lower
cusp crossing (non-planetary) model has $t_E = 147.3\,$days, $t_0 = 4042.5$,
$u_{\rm min} = -0.434$, $\theta = 6.044$, $\epsilon = 0.162$,
$t_\ast = 0.0519\,$days, and $R_M = 22.74$. These are the models
shown in Figure~\ref{fig-lc_bin1nopl}.
This figure indicates that the best fit model
is favored at $3885 < t < 3905$ because the non-planetary model
predicts higher magnification in the wings of the peak. But, in the
beginning of the 2007 season at $ 4135 < t < 4200$, the non-planetary,
upper or lower cusp model is favored in the MOA data, but not the OGLE data.
If these data are excluded, then the difference between the best fit model
(planetary) model increases to $\Delta\chi^2 = 73.4$. We regard this,
and all other, non-planetary models as being excluded by the data.

\section{Short Event Source Star Colors}
\label{sec-src_col}

In this Appendix (\ref{sec-src_col}), we compare the photometry of the short events
of \citet{sumi11} plus the binary events presented in this paper to 
the OGLE-III photometry catalog in order to determine calibrated
source magnitudes and colors.
\citet{gould_col} have argued that color of microlensed
source stars observed by both MOA and OGLE can be determined
from the differences between the MOA-Red passband and the OGLE-$I$
passband, which is similar to the standard Kron/Cousins $I$-band.
Since the MOA-Red band is essentially equivalent to the  sum of the
standard Kron/Cousins $R $ and $I$-bands, the color difference between
the MOA-Red  and OGLE-$I$ is rather small, and this implies that
the source brightness must be measured quite precisely in both passbands
in order to obtain a $V-I$ color of reasonable precision. This is an
important difficulty with the present analysis because the events
presented here and in \citet{sumi11} have brief periods of magnification
and because the OGLE survey involved is the relatively low-cadence
OGLE-III survey. The higher cadence of the current OGLE-IV survey
will help to alleviate this problem. As discussed in Section~\ref{sec-prop},
it was the low precision of the OGLE-III source flux measurement
that caused this method to fail for MOA-bin-1. This is also the reason
why this method fails for events MOA-ip-6, and 8 from \citet{sumi11}.
This method also cannot be used for MOA-ip-1, 5, and 10, because there
is no OGLE-III data during these events. Thus, we can only present
color measurements for the MOA-ip-2, 3, 4, 7 and 9 source stars in this
paper. For MOA-bin-1, MOA-bin-3 and the remaining \citet{sumi11} events,
we determine the transformation between the MOA and OGLE-III 
magnitude systems, but we can only estimate the source color
with a comparison to HST photometry. Event MOA-bin-3
does not lie in any OGLE-III field, so it is not included in this analysis.

As discussed in Section~\ref{sec-data}, the state of the art method
for crowded field photometry of time-varying sources is difference
image analysis (DIA) photometry. However, this method does not provide photometry
for the bright, relatively isolated stars that are needed to determine the
transformations between the passbands used for the observations
and standard passbands. Photometry of these stars must be done
with a point-spread function (PSF) fitting routine, such as DoPHOT
\citep{dophot}. As explained by \citet{gould_col}, there are substantial
differences in the way photometry is done with the DIA
and PSF fitting methods, and so there is an additional step to
determine the magnitude offset between the PSF fitting and DIA
photometry of the same images. 
This was done comparing the PSF fitting photometry of
apparently isolated stars to photometry done using an internal DIA
PSF fitting photometry algorithm. This later algorithm has no way to
handle blending. So, if too few sufficiently isolated stars can be found, it will be
impossible to get accurate photometry of the field stars using the
DIA fitting method. Of course, this entire method also depends on
PSF fitting photometry for the comparison of bright stars, but 
this is the situation that PSF fitting photometry codes were designed to handle:
the photometry of stars in  relatively crowded stellar fields. So, this
attempt to use the DIA PSF fitting routine on unsubtracted images
could be a major weakness of the \citet{gould_col} method.

We introduce a new method that does away with the need to find
stars that are so isolated that good photometry can be done
with the DIA PSF fitting method, which is designed for uncrowded,
subtracted frames. The PSF fitting method for a crowded field
PSF-fitting photometry code, like DoPHOT \citep{dophot} or
SoDoPHOT \citep{bennett-sod} should also work for uncrowded
difference images, so this is what we do.
We use the MOA DIA pipeline
\citep{bond01} for the difference image photometry and use both the
registered, unsubstracted frames and difference images from the
MOA pipeline as input for a custom version of SoDoPHOT.
The MOA DIA pipeline first generates registered
frames, which are transformed (and resampled) to have the same
astrometry as the reference image. 
Then the reference image is convolved
to the seeing of the registered image, and it is subtracted from the
registered image to give the difference image. The result is that
we have two sets of images, the registered images and the difference
images, with the same photometric scaling and the same PSF shape.
The registered frame has all the information of the stellar brightnesses 
from the original image, while the difference images have only the 
difference in flux from the reference frame. 

We then feed both the registered and difference images into our custom
version of SoDoPHOT. The registered frames are processed as normal
images, but the PSF model from each
registered frame is also used to do photometry on the target star in
the difference image. The result is a normal time series of SoDoPHOT
photometry from the registered frames plus photometry from the
difference images using the SoDoPHOT PSF model on the same photometric
scale as the as the PSF fitting photometry from SoDoPHOT. Thus, we
automatically have PSF and DIA photometry on the same photometric scale.
The SoDoPHOT generated DIA photometry may be slightly worse than the 
DIA photometry generated directly from the MOA pipeline, but in the 
examples presented here, we have found that both types of
photometry in these difference images were nearly identical.
The light curves presented in this paper use the original MOA pipeline
DIA photometry, while the color measurements discussed below use
SoDoPHOT PSF version of this photometry.

Now, in order to determine the color of the source stars for 5 of the 10
microlensing events with $t_E < 2\,$days from \citet{sumi11}, we
must consider three types of photometry:
\begin{itemize}
\item The OGLE-III catalog photometry, transformed to the standard Kron/Cousins 
$I$-band and the Johnson $V$-band (denoted by $I$ and $V$)
\citep{ogle3-phot}.
\item MOA-II photometry in the MOA-Red banded, denoted by $R_M$.
\item OGLE-III photometry in the native OGLE $I$-band, denoted by $I_O$.
\end{itemize}

The $R_M$ and $I_O$ magnitudes of the source ($R_{Ms}$ and $I_{Os}$)
are determined from the microlensing light curve fits to the joint
MOA and OGLE data sets. These magnitudes are indicated
in Table~\ref{tab-src_col}. Note that the error bars in the source magnitudes
in this table assume that the parameters of the microlensing model are
fixed at the best fit values, but they do not include any effect of
light curve model uncertainties. 

As discussed in \citet{ogle3-phot}, the relationship between the native $I_O$ passband
is non-linear due some transparency in the OGLE-III $I_O$-band filter beyond the
wavelength range of the standard Cousins $I$-band. This relation is
\begin{equation}
I = I_O - d + \left(b_{\rm CCD} + g\right)\left(V-I\right) + h\left(V-I\right)^2 \ ,
\label{eq-ogle_cal}
\end{equation}
where $d = 0.033918$, $g = 0.016361$ and $h = 0.004167$. The parameter 
$b_{\rm CCD}\sim 0.04$ is necessary because of the slight variations in 
sensitivity as a function of wavelength for the eight different CCD chips in the
OGLE-III camera \citep{ogle-pipeline}.

In order to extract standard color information from the $R_M - I_O$ data available
from the light curves we need an expression relating $R_M$ to the standard
Johnson-$V$ and Cousins-$I$ bands. This is accomplished with a comparison
of SoDoPHOT photometry of the field of each MOA event to the same field
in the OGLE-III catalog. As discussed by \citet{gould_col}, it is important
to ensure that comparison stars do not have close neighbors in the OGLE catalog
because the MOA images do not have seeing as good as the images used to
construct the OGLE-III photometry catalog. 

For the purpose of this analysis, we limit this MOA-OGLE color comparison to stars
with colors slightly bluer than the top of the bulge main sequence to 
a few $\times 0.1\,$mag redder than the red clump, as this covers the range of
plausible source star colors. This allows us to use a linear relation of the form
\begin{equation}
R_M = I + f\left(V-I\right) \ ,
\label{eq-moa_cal}
\end{equation}
where the coefficient $f\sim 0.2$ depends on the particular field. 
(This field dependence is probably due to the width of the $R_M$ passband
and the wide variation of interstellar extinction in the foreground of the 
Galactic bulge fields.

\begin{figure}
\plotone{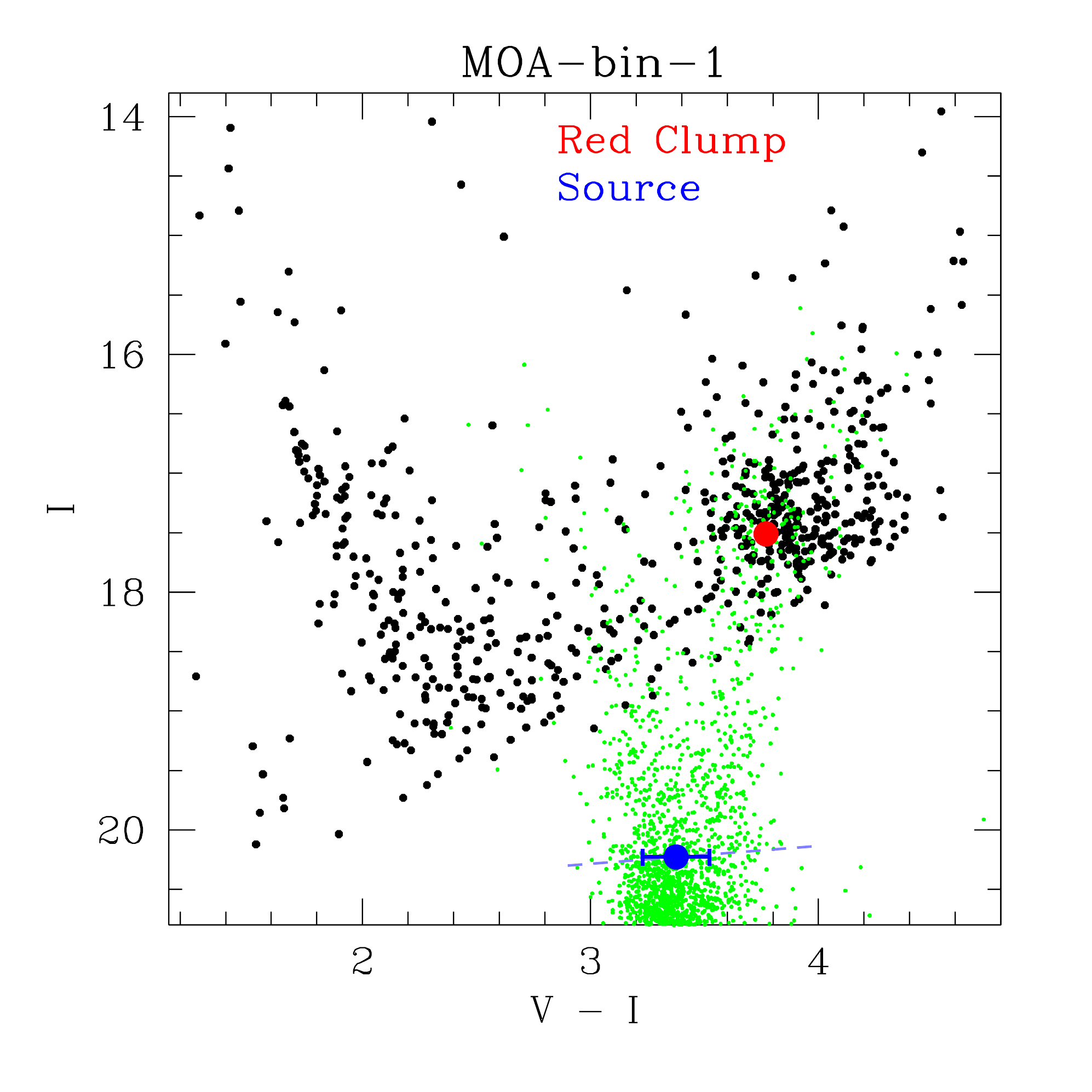}
\caption{The color magnitude diagram (CMD) of the stars in the OGLE-III catlog
within $60^{\prime\prime}$ of MOA-bin-1 is shown as black dots, while the
green dots show the HST CMD of \citet{holtzman98}, transformed to the
same extinction (and extinction dispersion) as the field of MOA-bin-1.
The red dot shows the centroid of the red clump giant distribution. The
light blue dashed line shows the constraint based on the MOA-red
source brightness of $R_{Ms} = 20.73$, and
the blue dot (and errorrbars) shows the source brightness 
$I_s = 20.23\pm 0.02 $ and color $(V-I)_s = 3.38 \pm 0.15$ estimated
from a comparison with the HST CMD.
\label{fig-cmd_bin1}}
\end{figure}

\begin{figure}
\plotone{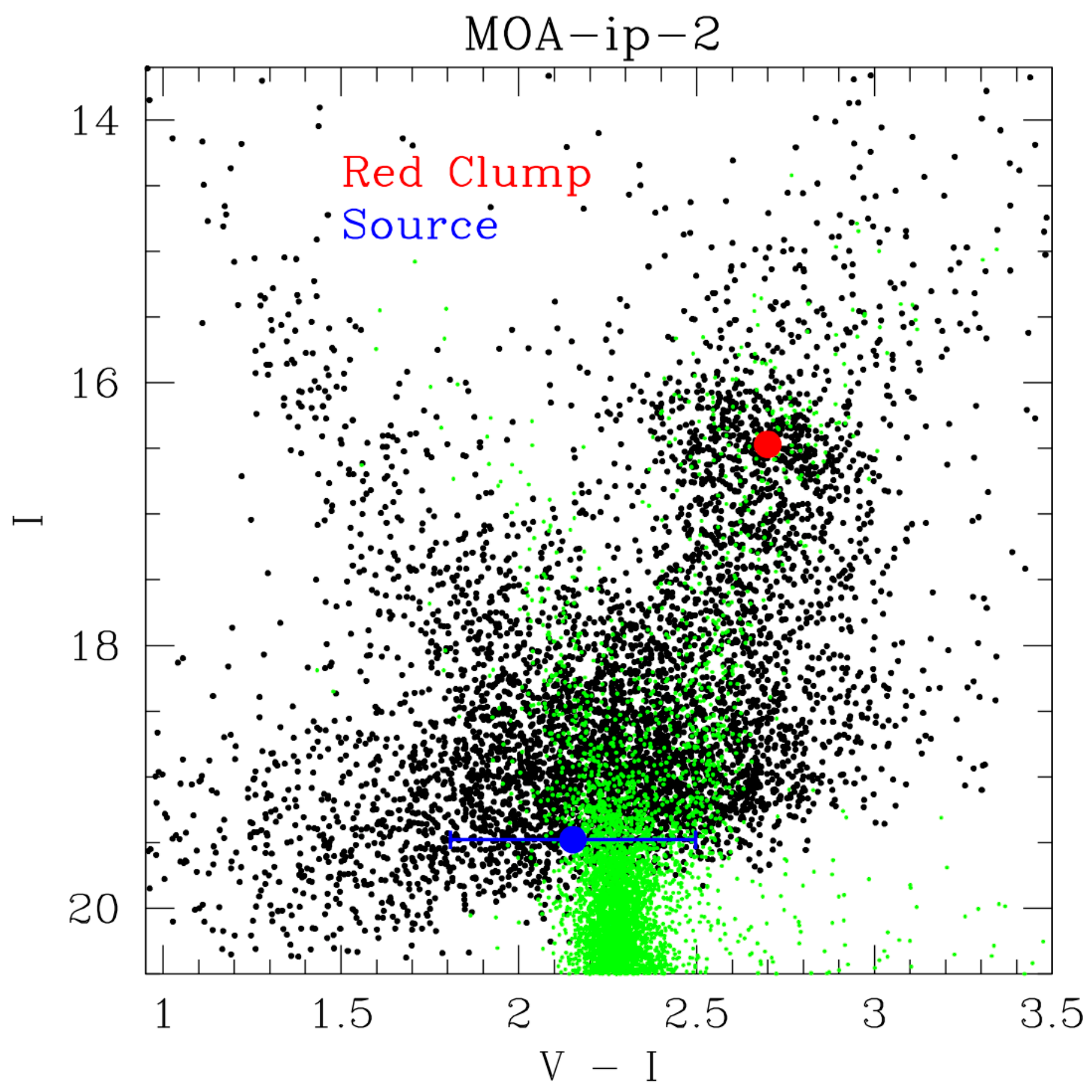}
\caption{The CMD of the stars in the OGLE-III catlog
within $120^{\prime\prime}$ of MOA-ip-2 is shown as black dots, while the
green dots show the HST CMD of \citet{holtzman98}, transformed to the
same extinction as the field of MOA-ip-2. The centroid of the red clump
is shown as a red dot, and the $I$ magnitude and color
of the source star are shown in blue.
\label{fig-cmd_ip2}}
\end{figure}

\begin{figure}
\plotone{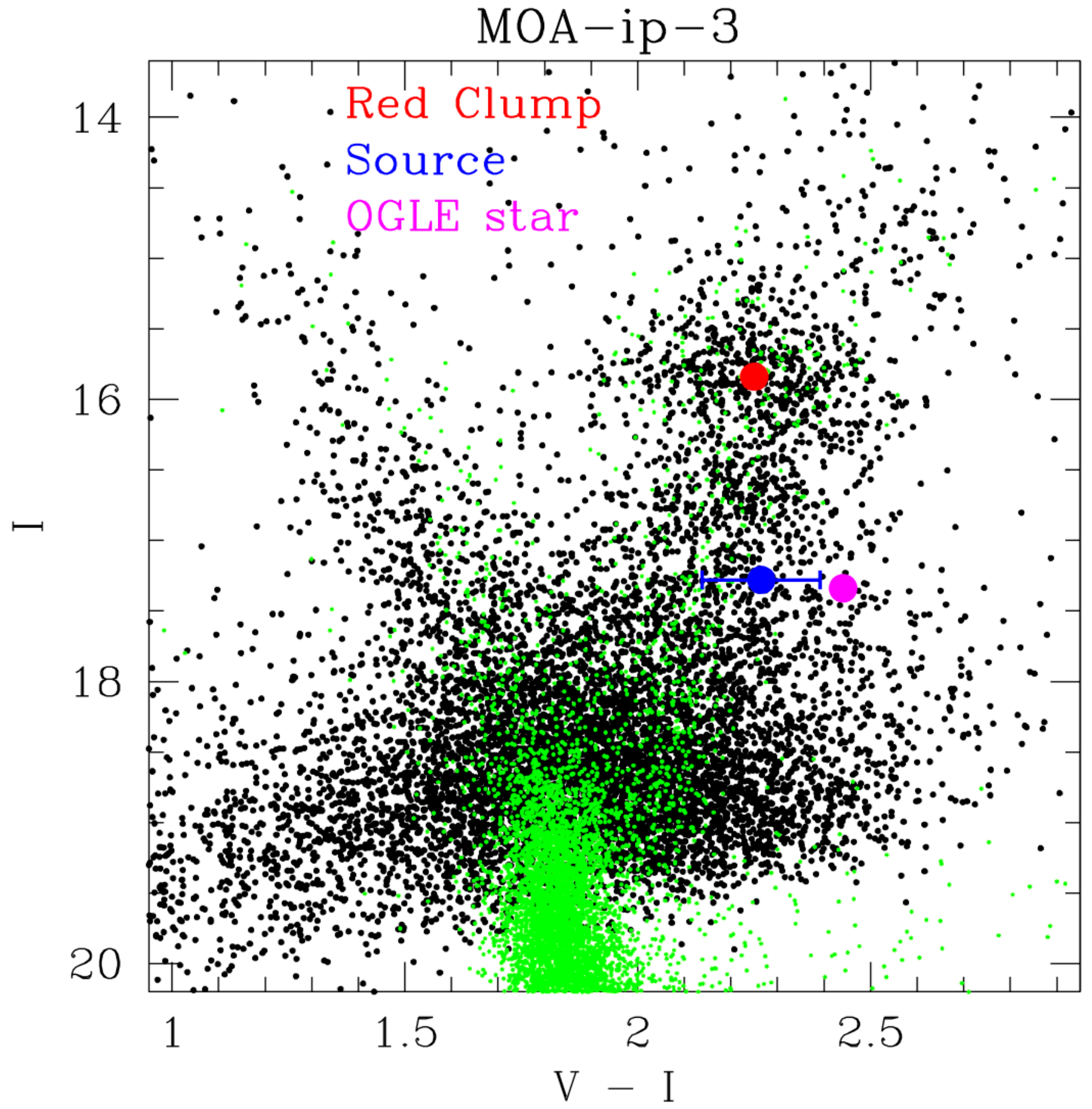}
\caption{The CMD of the stars in the OGLE-III catlog
within $120^{\prime\prime}$ of MOA-ip-3 are
shown as black dots, and the
green dots show the HST CMD of \citet{holtzman98}, transformed to the
same extinction as the field of MOA-ip-3. The centroid of the red clump
is shown as a red dot; $I$ magnitude and color
of the source star are shown in blue, and the 
with the color and magnitude of the star in the OGLE-III catalog
located at the position of the event is shown as a magenta dot.
\label{fig-cmd_ip3}}
\end{figure}

\begin{figure}
\plotone{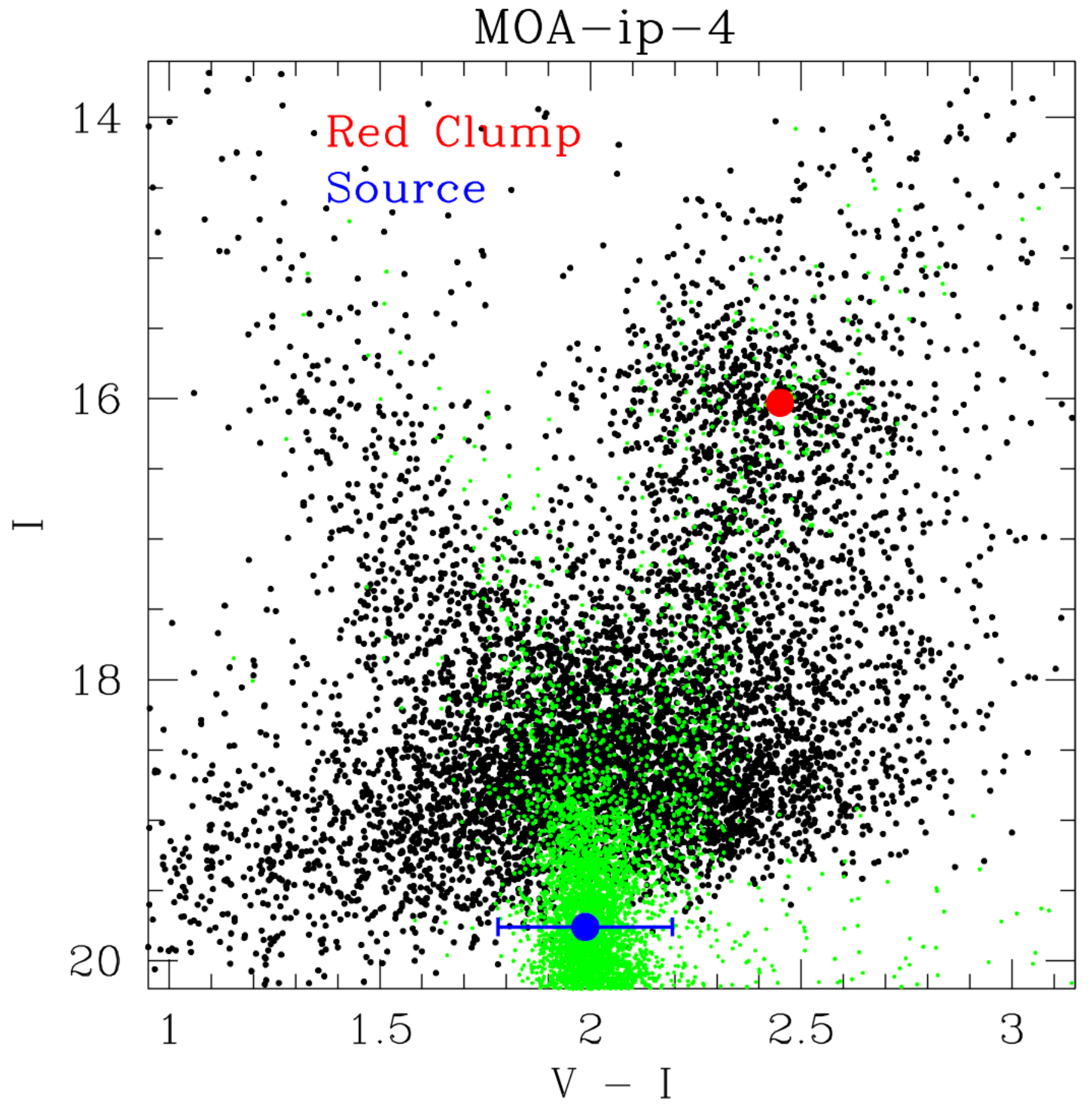}
\caption{The CMD of the stars in the OGLE-III catlog
within $120^{\prime\prime}$ of MOA-ip-4. The catalog
stars, HST stars, red clump centroid and source star are indicated
as in Figure~\ref{fig-cmd_ip2}.
\label{fig-cmd_ip4}}
\end{figure}

\begin{figure}
\plotone{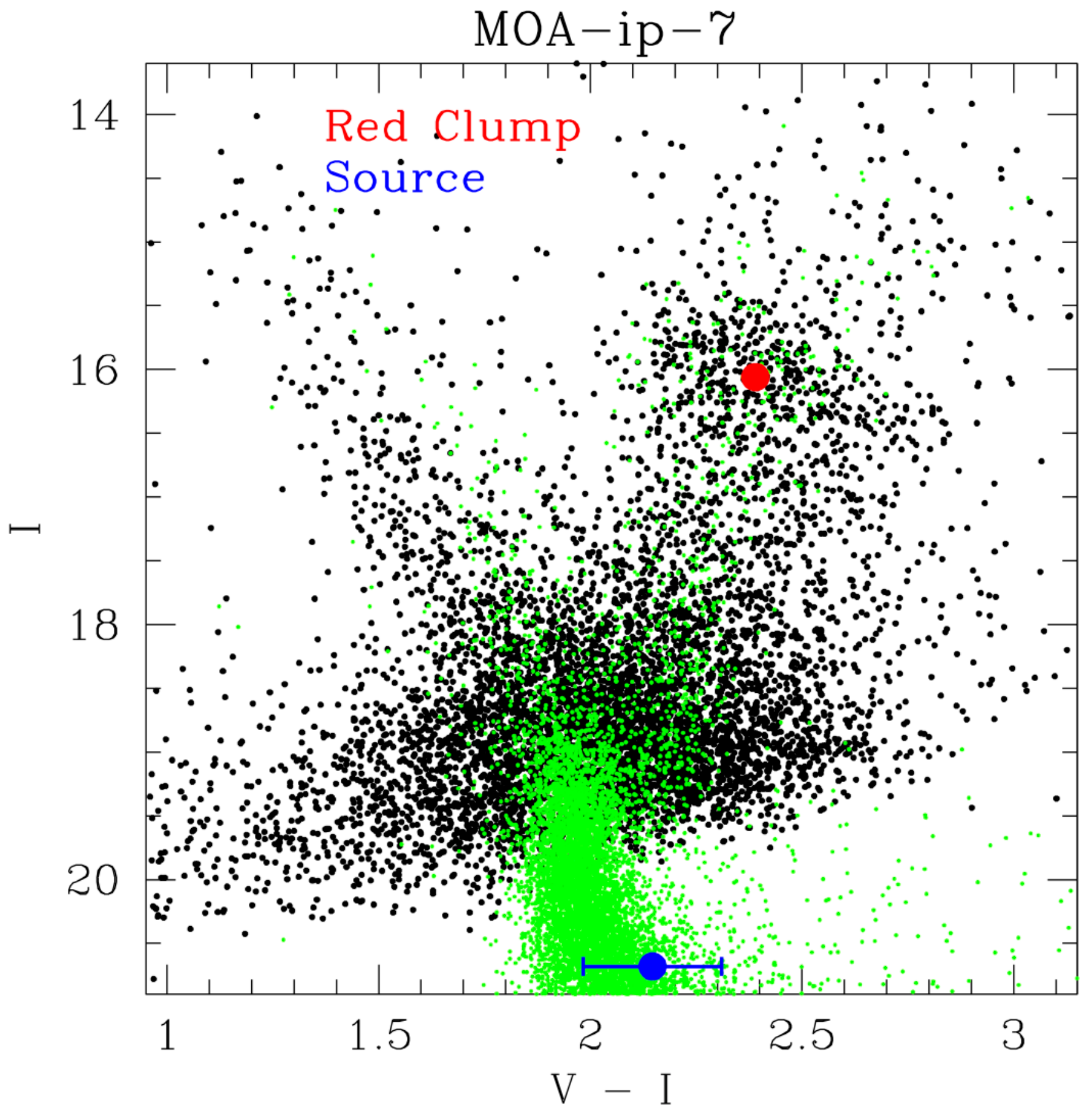}
\caption{The CMD of the stars in the OGLE-III catlog
within $120^{\prime\prime}$ of MOA-ip-7. The catalog
stars, HST stars, red clump centroid and source star are indicated
as in Figure~\ref{fig-cmd_ip2}.
\label{fig-cmd_ip7}}
\end{figure}

\begin{figure}
\plotone{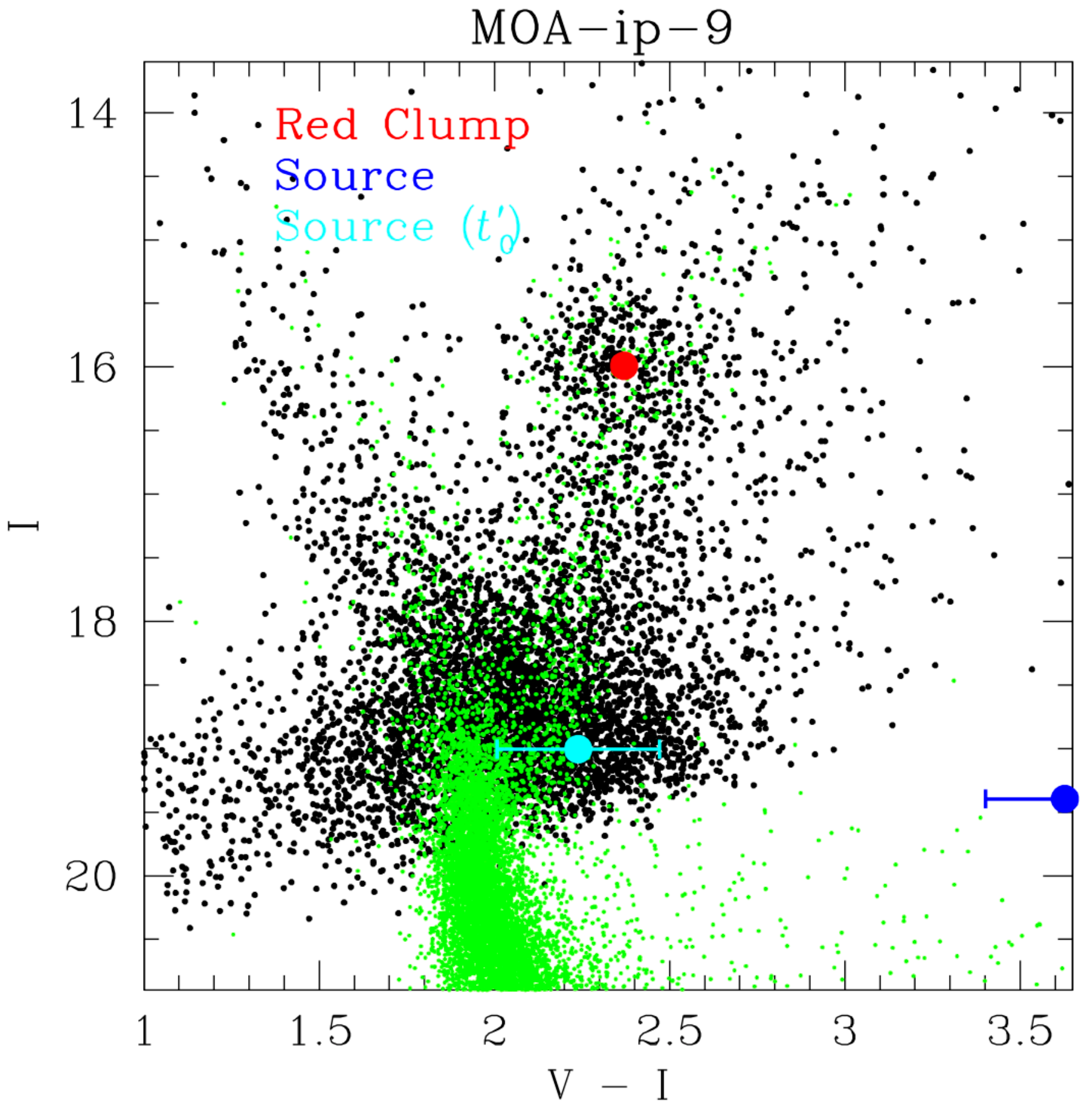}
\caption{The CMD of the stars in the OGLE-III catlog
within $120^{\prime\prime}$ of MOA-ip-9. The catalog
stars, HST stars, red clump centroid and source star are indicated
as in Figure~\ref{fig-cmd_ip2}. The cyan dot indicates the location
of the source with a different $t_0$, which changes the color estimate
to one is consistent with a bulge sub-giant source. This solution is
disfavored by $\Delta\chi^2 = 2.36$.
\label{fig-cmd_ip9}}
\end{figure}

The results of this analysis, as well as the adopted $b_{\rm CCD}$ and $f$ 
parameters given in Table~\ref{tab-src_col} for events MOA-ip-2, 3, 5, 7 and 9. 
Figures~\ref{fig-cmd_ip2}-\ref{fig-cmd_ip9} display CMDs for all the stars
within 2 arc min of each event, as well as the locations of the adopted
red clump centroid and the magnitude and color of the source star, which
has been determined by this analysis. As in Figure~\ref{fig-cmd_bin1}, we
also display HST bulge CMD from \citet{holtzman98} as green points, and we
have aligned the HST CMD to the OGLE-III CMD by matching the
red clump centroid appropriate for the Holtzman Baade's Window field to
the observed red clump position in the field of each event. But, unlike
Figure~\ref{fig-cmd_bin1}, we do not broaden the CMD in an attempt to
simulate the excess differential extinction in these fields compared to 
the Baade's window.

The results for MOA-ip-2, 4, and 7 are similar. In each case,
we find a source brightness and color consistent with a main sequence
source star. For MOA-ip-2, the source is close to the main sequence 
turn-off, while the source for MOA-ip-4 and 7 are fainter, with the MOA-ip-7
source about 1.5 mag below the turn-off. All three of these sources are 
basically below the crowding limit in the OGLE-III catalog, so we cannot 
expect to identify the sources in images with $\sim 1^{\prime\prime}$ seeing.

The one source that does appear to be identified in the OGLE-III catalog
is MOA-ip-3, and we indicate the CMD position of the apparent source
as a magenta dot in Figure~\ref{fig-cmd_ip3}. The microlensing fit
indicates that the source is in the middle of the sub-giant branch of the
CMD about half-way between the base and the red clump. It is not surprising
that the source appears to be on the sub-giant branch, while the OGLE star
at this location is considerably redder than any of the subgiants on the HST
CMD. This is fairly typical, as these bulge fields are quite crowded when observed
in $\sim 1^{\prime\prime}$ seeing, so the CMDs tend to have significant
errors due to blending, particularly at faint magnitudes. Blending is much less of
a problem with the $\simlt 0.1^{\prime\prime}$ seeing of HST. For two sources to
be simultaneously microlensed by the same lens requires an angular
separation of $\simlt 1\,$mas, which is virtually impossible unless 
the source is a gravitationally bound binary star system.

The main reason for the separation between the source and the OGLE-star position
in  Figure~\ref{fig-cmd_ip3} is likely to be blending related errors in the OGLE-III
CMD. The source star is slightly brighter than the 
apparent star in the OGLE catalog and has a color of $V_s-I_s = 2.27\pm 0.13$,
which is 1.3-$\sigma$ bluer than the OGLE-III catalog color 
of $(V-I)_{\rm cat} = 2.44$. Both of these effects are likely to due to crowding.
This is why the OGLE catalog position in the CMD is not sparsely populated with 
OGLE stars, while HST stars are absent from this part of the CMD. Blending
can also produce situations in which appears to be blended with a target star
of negative brightness.
Most of the bulge main sequence stars are not resolved in the OGLE and MOA
images, and so they contribute to the apparent background light. If the source
happens to be located near a local minima of the unresolved star background, 
it will appear to be blended with a star of negative brightness.

The one event that seems to give a very strange source color is MOA-ip-9,
where the best fit light curve model implies a color of $V_s-I_s = 3.63\pm0.23$,
which is more than a magnitude redder than the red clump. However, the
light curve shown in Figure~S1 of \citet{sumi11} shows that the MOA data
cover the falling part of the light curve, while OGLE has only a single
significantly magnified observation on the rising portion of the light curve.
This implies that the implied color is likely to be strongly correlated
with the model parameters, and $t_0$, in particular. It we fix 
$t_0 = t_0^\prime = 2453910.7650\,{\rm HJD}$, we find that the
fit $\chi^2$ increases by $\Delta\chi^2 = 2.36$, but the implied
source color drops to $V_s-I_s = 2.24\pm0.23$, as shown in 
cyan in Figure~\ref{fig-cmd_ip9} and listed in Table~\ref{tab-src_col}.
This matches the color of the base of the giant branch in this field
and is consistent with the color of the main sequence turn-off as well.
So, this is consistent with the microlensing model at $\sim 1.5\,\sigma$.
It also indicates that for this event, and others like it, there is a substantial
source color uncertainty associated with the microlensing model parameters.
Event MOA-ip-2 has similar light curve coverage, so it probably also
falls in this category.

For the remaining three events, MOA-ip-3, 4, and 7, these color estimates
do tend to confirm the microlensing models.

Finally, for the events in OGLE-III fields that do not have sufficient 
OGLE-III photometry while they are magnified, we estimate the
colors following the HST comparison method used for MOA-bin-1 as
described in the main text. The results for these stars are indicated
in Table~\ref{tab-src_mag}.

\clearpage


\clearpage

\begin{deluxetable}{ccccccc}
\tablecaption{Event Names and Coordinates
                         \label{tab-coords} }
\tablewidth{0pt}
\tablehead{
\colhead{Event}  &\colhead{Field}  &  \colhead{Classification} &
\colhead{RA} & \colhead{DEC} & \colhead{$l$} & \colhead{$b$}
}  

\startdata

MOA-bin-1 & gb5 & planetary & 17:51:10.20 & -29:47:38.3 &  -0.111 & -1.478\\
MOA-bin-2 & gb19 & brown dwarf & 18:15:20.36 & -25:24:43.2 & 6.342 & -3.999\\
MOA-bin-3 & gb5 & brown dwarf & 17:51:04.40 & -28:46:48.9 & 0.750 & -0.942 \\
flare & gb9 & foreground flare & 18:02:17.59 & -29:01:20.5 & 1.772 & -3.188 \\
\enddata

\tablecomments{The locations of the events presented in this paper in
equatorial (J2000) and Galactic ($l$ and $b$) coordinates. }

\end{deluxetable}

\begin{deluxetable}{ccccc}
\tablecaption{Non-Linear Model Parameters for MOA-bin-1
                         \label{tab-mparam} }
\tablewidth{0pt}
\tablehead{
\colhead{parameter}  & \colhead{units} &
\colhead{best fit} & \colhead{outer cusp model} & \colhead{MCMC range} 
}  

\startdata

$t_E$ & days & 31.769 & 31.233 & $31.2\pm 4.3$ \\
$t_0$ & ${\rm HJD}-2450000$ & 3892.9777 & 3894.8444 & $3893.8\pm 1.8$ \\
$u_{\rm min}$ & & 1.6197 & 1.6177 & $1.62\pm 0.05$ \\
$s$ & & 2.1272 &2.0795 & $2.10 \pm 0.05$ \\
$\theta$ & radians & 1.5947 & 1.5582 & $ 1.578\pm 0.035$ \\
$\epsilon$ & & $4.560\times 10^{-3}$ & $4.194\times 10^{-3}$ & $4.8{+1.6\atop -1.2}\times 10^{-3}$ \\
$t_\ast$ & days & 0.05089 & 0.04999 & $0.0500\pm 0.0010$ \\
$R_M$ & mag & 20.73 & 20.75 & $20.76\pm 0.08$ \\
fit $\chi^2$ & for 7223 d.o.f. & 7230.637 & 7231.119 & \\

\enddata


\end{deluxetable}

\begin{deluxetable}{ccccc}
\tablecaption{Non-Linear Model Parameters for MOA-bin-2, bin-3, and Flare Star event
                         \label{tab-mparams} }
\tablewidth{0pt}
\tablehead{
\colhead{parameter}  & \colhead{units} &
\colhead{MOA-bin-2} & \colhead{MOA-bin-3} & \colhead{Flare} 
}  

\startdata

$t_E$ & days & 15.291 & 10.101 & 0.911  \\
$t_0$ & ${\rm HJD}-2450000$ & 3874.6568 & 4224.9102 & 4182.8797 \\
$u_{\rm min}$ & & 0.4140 & 1.0191 & 1.3708 \\
$s$ & & 1.1629 & 0.5567 & 2.0355 \\
$\theta$ & radians & 2.2853 & -0.1350 & 1.0804 \\
$\epsilon$ & & 0.0473 & 0.0946 & 0.0104 \\
$t_\ast$ & days & 0.0202 & 0.0250 & 0.00952 \\
data points & & 1599 & 13667 & 6499 \\
fit $\chi^2$ &  & 1585.951 & 13657.543 & 6471.159  \\

\enddata


\end{deluxetable}

\begin{deluxetable}{ccccccc}
\tablecaption{Physical Parameters for MOA-bin-1L
                         \label{tab-pparam} }
\tablewidth{0pt}
\tablehead{
 & & \multicolumn{5}{c} {parameter limits} \\
\colhead{parameter}  & \colhead{units} &
\colhead{$-2\sigma$} & \colhead{$-1\sigma$} &  \colhead{\bf median} &
\colhead{$+1\sigma$} & \colhead{$+2\sigma$} 
}  

\startdata

$D_L$ & kpc  & 1.0 & 3.2 & {\bf 5.1} & 6.3 & 7.3 \\
$M_\ast$ & $\msun $  & 0.09 & 0.34 & {\bf 0.75} & 1.08 & 1.26 \\
$m_p$ & $M_{\rm Jup}$   & 0.4 & 1.6 & {\bf 3.7} & 5.8 & 8.3 \\
$a$ & AU & 2.1 & 5.6 & {\bf 8.3} & 12.8 & 29.7 \\
$I_L$ & mag & 26.55 & 24.77 & {\bf 22.59} & 20.90 & 20.04 \\
\enddata

\tablecomments{The planetary system distance, $D_L$, host star mass, $M_*$, 
planet mass, $m_p$, the semi-major axis, $a$, and host star brightness $I_L$. }

\end{deluxetable}

\begin{deluxetable}{ccccc}
\tablecaption{Wide Orbit Host Detection Probability for $dN/ds \propto s^n$
                         \label{tab-hostprob} }
\tablewidth{0pt}
\tablehead{
 & \multicolumn{4}{c} {Probability per Model} \\
\colhead{$N$}  & \colhead{$n=-1$} & \colhead{$n=-1$} & \colhead{$n=0$} &  \colhead{$n=0$ } \\
  & \colhead{$s_1 = 60$} & \colhead{$s_1 = 135$} & \colhead{$s_1 = 18$} &  \colhead{$s_1 = 21$ }
}  

\startdata

0 & 0.0500 & 0.1009 & 0.0482 & 0.1026 \\
1 & 0.1950 & 0.2782 & 0.2813 & 0.3483 \\
2 & 0.3044 & 0.3187 & 0.3636 & 0.3392 \\
3 & 0.2600 & 0.2007 & 0.2170 & 0.1589 \\
4 & 0.1342 & 0.0782 & 0.0729 & 0.0430 \\
5 & 0.0451 & 0.0198 & 0.0148 & 0.0072 \\
6 & 0.0097 & 0.0032 & 0.0019 & 0.0007 \\
7 & 0.0013 & 0.0004 & 0.0002 & 0\ \ \ \ \ \ \ \ \\
8 & 0.0001 & 0\ \ \ \ \ \ \ \  & 0\ \ \ \ \ \ \ \ &0\ \ \ \ \ \ \ \ \\
9 & 0\ \ \ \ \ \ \ \ & 0\ \ \ \ \ \ \ \  & 0\ \ \ \ \ \ \ \ &0\ \ \ \ \ \ \ \ \\
10 & 0\ \ \ \ \ \ \ \ & 0\ \ \ \ \ \ \ \  & 0\ \ \ \ \ \ \ \ &0\ \ \ \ \ \ \ \ \\
\enddata

\tablecomments{Assumes a non-zero wide orbit probability explaining
all the isolated planets of the \citet{sumi11} sample with
$dN/ds  \propto s^n$ for $s_0 < s < s_1$ and $s_0 = 2$. }

\end{deluxetable}

\begin{deluxetable}{ccccccc}
\tablenum{B1}
\tablecaption{Isolated Planet Event Source Color Measurements
                         \label{tab-src_col} }
\tablewidth{0pt}
\tablehead{
\colhead{Event}  & \colhead{$R_{Ms}$} & \colhead{$I_{Os}$} & \colhead{$b_{\rm CCD}$} &
\colhead{$f$} & \colhead{$I_s$} &  \colhead{$V_s-I_s$} 
}  

\startdata

MOA-ip-2 & $19.882\pm 0.019$ &$19.295\pm 0.087$ & 0.044 & 0.1877 & $19.48\pm 0.07$ & $2.15\pm 0.34$ \\
MOA-ip-3 & $17.746\pm 0.004$ &$17.088\pm 0.027$ & 0.044 & 0.2060 & $17.28\pm 0.03$ & $2.27\pm 0.13$ \\
MOA-ip-4 & $20.077\pm 0.010$ &$19.593\pm 0.053$ & 0.042 & 0.1599 & $19.76\pm 0.03$ & $1.99\pm 0.21$ \\
MOA-ip-7 & $21.092\pm 0.010$ &$20.503\pm 0.042$ & 0.042 & 0.1912 & $20.68\pm 0.03$ & $2.15\pm 0.16$ \\
MOA-ip-9 & $20.052\pm 0.018$ &$19.095\pm 0.046$ & 0.042 & 0.1809 & $19.40\pm 0.04$ & $3.63\pm 0.23$ \\
MOA-ip-9($t_0^{\prime})$ & 
                 $19.410\pm 0.018$ &$18.819\pm 0.046$ & 0.042 & 0.1809 & $19.00\pm 0.05$ & $2.24\pm 0.23$ \\
\enddata

\tablecomments{Source magnitudes, $R_{Ms}$ and $I_{Os}$, color transformation
parameters, $b_{\rm CCD}$ and $f$, and estimated source magnitude and color
for 5 of the 10 isolated planet events from \citet{sumi11}. }
\end{deluxetable}

\begin{deluxetable}{ccccc}
\tablenum{B2}
\tablecaption{Host Star Magnitudes and Color Estimates
                         \label{tab-src_mag} }
\tablewidth{0pt}
\tablehead{
\colhead{Event}  & \colhead{$R_{Ms}$} &
\colhead{$f$} & \colhead{$I_s$} &  \colhead{$V_s-I_s$} 
}  

\startdata

MOA-bin-1 & 20.73 & 0.1485 & 20.23 & 3.38 \\
MOA-bin-2 & 21.59 & 0.1830 & 21.21 & 2.08 \\
MOA-ip-1 & 20.45 & 0.1730 & 20.11 & 1.96 \\
MOA-ip-5 & 20.23 & 0.2236 & 19.79 & 1.97 \\
MOA-ip-6 & 19.18 & 0.1907 & 18.81 & 1.94 \\
MOA-ip-8 & 19.61 & 0.2122 & 19.27 & 1.62 \\
MOA-ip-10 &19.71 & 0.2053 & 19.42 & 1.38 \\
\enddata

\tablecomments{Source magnitude, $R_{Ms}$ color transformation
parameters, $f$, and estimated source magnitude and color
for 2 of the 3 short binary events and 
5 of the 10 isolated planet events from \citet{sumi11}. The
color is based on a comparison to an HST luminosity function \citep{holtzman98}. }
\end{deluxetable}

\end{document}